%% file: main_iclr.tex
\documentclass{article} % For LaTeX2e

\usepackage{iclr2027_conference,times}

\input{math_commands.tex}

\usepackage{hyperref}
\usepackage{url}
\usepackage{booktabs}       % professional-quality tables
\usepackage{amsfonts}       % blackboard math symbols
\usepackage{nicefrac}       % compact symbols for 1/2, etc.
\usepackage{microtype}      % microtypography
\usepackage{xcolor}         % colors
\usepackage{amsmath} 
\usepackage{amssymb}
\usepackage{bbm}
\usepackage{tabularx}
\usepackage{graphicx}
\usepackage{multirow} 
\usepackage{caption}
\usepackage{enumitem}
\usepackage{placeins}
\usepackage{wrapfig}
\usepackage{needspace}
\usepackage{xcolor}
\usepackage{comment}
\usepackage{algorithm}
\usepackage{algpseudocode}

\title{ProteoCraft: Protein Design as Proposal–Verification Search}

\author{%
\bfseries
Fang Wu$^{1}$\thanks{Equal contribution. Correspondence to
\texttt{fangwu97@stanford.edu}, \texttt{yejinc@stanford.edu}.}
\quad Hongbin Lin$^{2}$\footnotemark[1]
\quad Bingchen Duan$^{2}$\footnotemark[1]
\quad Yuchen Wang$^{1}$\footnotemark[1]
\quad Li Erran Li$^{3}$\footnotemark[1]
\\
\bfseries
Yanjie Huang$^{4}$\footnotemark[1]
\quad Mingxuan Cao$^{5}$
\quad Yijun Ma$^{6}$
\quad Yunxiang Mo$^{7}$
\quad Yanchao Li$^{8}$
\\
\bfseries
Kuan Pang$^{1}$
\quad Juangui Xu$^{9}$
\quad Hanqun Cao$^{10}$
\quad Dongwei Jiang$^{1}$
\quad Xinyue Xu$^{7}$
\\
\bfseries
Xiangru Tang$^{11}$
\quad Zehong Wang$^{6}$
\quad Jian Ma$^{3}$
\quad Kejun Ying$^{1}$
\quad Guanlue Li$^{12}$
\\
\bfseries
Mingran Cao$^{13}$
\quad Fanjin Wang$^{13}$
\quad Tianfan Fu$^{8}$
\quad Shuangjia Zheng$^{4}$
\\
\bfseries
Yao Shu$^{2}$
\quad Jure Leskovec$^{1}$
\quad Yejin Choi$^{1}$
\\[0.5em]
\normalfont
$^{1}$Stanford University
\quad $^{2}$HKUST (Guangzhou)
\quad $^{3}$Amazon
\quad $^{4}$SJTU
\\
$^{5}$University of Chicago
\quad $^{6}$University of Notre Dame
\quad $^{7}$HKUST
\\
$^{8}$Nanjing University
\quad $^{9}$CISPA
\quad $^{10}$CUHK
\quad $^{11}$Yale University
\\
$^{12}$Universit\"at Hamburg
\quad $^{13}$Lexicodon
}
\iclrfinalcopy % Uncomment for camera-ready version, but NOT for submission.
\begin{document}

\maketitle

\begin{abstract}
    Protein–protein interactions underpin nearly all biological processes.  Yet, designing proteins with desired interactions remains challenging due to the structural complexity governing molecular recognition. This study views protein design as a search optimization problem over a combinatorial sequence space under expensive structural evaluation. Toward this goal, we introduce a novel tree-based search framework, ProteoCraft, that employs approximate structural gradients from diffusion-based structure predictors as effective proposal policies to specify where the search should explore. Complete no-gradient inference of folding models separately assigns values to the resulting sequences, determining what is good. ProteoCraft allocates this verification budget through a coarse-to-fine paradigm: multi-site edits first explore distinct sequence basins, then verified promising states undergo strict single-site refinement. Across fourteen de novo binder-design settings spanning eight target proteins, five immune-relevant TCR-facing targets, and four localized CDR redesign tasks, ProteoCraft improves design quality and high-confidence yield. These improvements persist under independent AF2-IG and Rosetta-based interface evaluation. Prospective experimental validation further yielded an 82.9\% binding hit rate for PD-L1 miniproteins, with a best measured $K_D$ of 0.879\,nM, and confirmed binding of designed IFNAR2 VHH. Our code is available at~\url{https://github.com/smiles724/ProteoCraft_v1}.
\end{abstract}

\input{sec/1_intro}

\input{sec/3_method}

\input{sec/4_experiment}

\section{Conclusion}
ProteoCraft shows that approximate gradients from diffusion predictors need not serve as faithful value estimators to be useful for protein design. By using differentiable signals only to propose discrete sequence edits, and reserving complete predictor inference for candidate evaluation, ProteoCraft turns diffusion-based protein design into a proposal–verification search problem. A coarse-to-fine search then allocates expensive verifier calls between broad multi-site exploration and conservative single-site refinement.

\subsection*{AI Use Statement}

Generative AI tools were used to assist with language editing, literature discovery, and limited code development and debugging. All suggested references were verified by the authors against the original publications, and all AI-assisted code was manually reviewed and tested for correctness. Generative AI tools were not used to generate experimental data, fabricate results, or autonomously determine the scientific claims or conclusions of this work. The authors designed the methodology, conducted the experiments, analyzed the results, and verified all reported findings. We take full responsibility for the final content of this work, including any text, code, or other artifacts produced
with the assistance of generative AI.

\bibliography{cite}
\bibliographystyle{iclr2027_conference}

\appendix
\newpage 

\input{sec/2_related}
\input{sec/X_appendix}

\end{document}

%% file: math_commands.tex
\usepackage{amsmath,amsfonts,bm}

\def\eqref#1{equation~\ref{#1}}
\def\1{\bm{1}}

\DeclareMathAlphabet{\mathsfit}{\encodingdefault}{\sfdefault}{m}{sl}
\SetMathAlphabet{\mathsfit}{bold}{\encodingdefault}{\sfdefault}{bx}{n}

%% file: sec/1_intro.tex
\begin{figure}[ht]
    \centering
    \vspace{-1.5em}
    \includegraphics[width=0.9\linewidth]{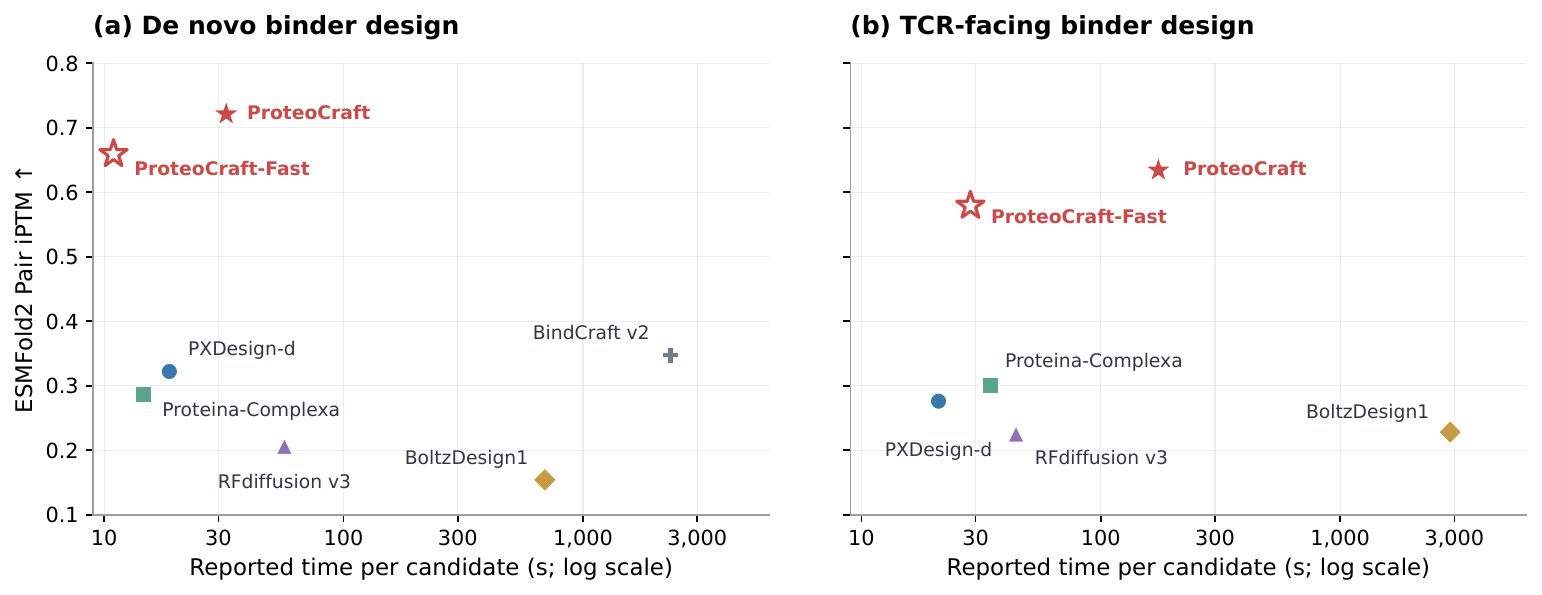}
    \vspace{-0.5em}
    \caption{Design quality versus computational cost.}
    \vspace{-1.5em}
    \label{fig:latency_performance}
\end{figure}

\section{Introduction}

Protein--protein interactions (PPIs) underlie many cellular processes, and the ability to design proteins that selectively engage a target has broad applications in therapeutics, diagnostics, and synthetic biology~\citep{zambaldi2024novo,cho2025protein}. Computational protein design offers a powerful alternative by enabling binders to be tailored to specific targets and epitopes. Early physics-based approaches, such as Rosetta~\citep{cao2022design,gainza2023novo}, designed binders through scaffold docking and side-chain optimization. However, these methods typically have low experimental success rates and therefore require screening a large number of candidates~\citep{bennett2023improving}. Their reliance on docking predefined scaffolds onto rigid target structures can also produce geometric or chemical incompatibilities at the interface, resulting in suboptimal binding or making certain epitopes difficult to target.

Recent advances in deep learning (DL) have transformed biomolecular modeling and design. AlphaFold2/3 (AF2/3)~\citep{jumper2021highly,evans2021protein,abramson2024accurate}, for example, demonstrated remarkable accuracy in predicting monomer and complex structures. AF-based filtering substantially improves binder-design success rates by identifying candidates predicted to form plausible complexes with their targets~\citep{liu2026protdbench,team2025pxdesign}. DL has also enabled the de novo generation of protein binders. A widely used pipeline combines RFdiffusion for backbone generation~\citep{watson2023novo} with ProteinMPNN~\citep{dauparas2022robust} for sequence design, achieving markedly higher experimental success rates. Nevertheless, it separates backbone generation from sequence design without explicitly modeling full atomic interactions, leaving AF filtering to bridge the gap between structural generation and functional interface design.
Building on the success of AF-based filtering, another hallucination-based line, such as BindCraft~\citep{pacesa2024bindcraft}, demonstrated that structure predictors can also serve as differentiable design engines. By backpropagating through AF2, they directly optimize binder sequences, producing high-affinity designs with substantially reduced experimental screening. This marked a shift from generate-then-filter pipelines toward prediction-guided optimization. However, direct gradient optimization can be myopic and exploit imperfections in the predictive model, motivating search-based approaches that combine gradient-guided proposals with a balanced exploration-exploitation trade-off.

% In parallel, all-atom generative models—including Latent-X~\citep{team2025latent}, UniMoMo~\citep{kong2025unimomo}, AnewOmni~\citep{kong2026programming}, BoltzGen~\citep{stark2025boltzgen}, and Chai-2~\citep{chai2025zero}—directly generate binder sequences and atomic structures. Although these models improve atomic-level interface modelling, their iterative generative processes make direct gradient-based sequence optimization substantially more expensive than with feed-forward predictors. 

In this paper, we formulate protein design as a combinatorial search problem, where nodes represent discrete sequences, edits define transitions between them, and structure predictors serve as verification oracles. Our central idea is to use gradient-based hallucination as a source of search directions, while using structural verification to decide which sequences to explore further. To this end, we introduce ProteoCraft, a tree-based search framework based on ESMFold2~\citep{candido2026language}. Specifically, gradients through the terminal diffusion step induce priors over residue substitutions, which guide a coarse-to-fine search from coordinated multi-site exploration to single-site refinement. Verification therefore becomes part of the search process rather than only a final filter. Each proposed sequence is independently evaluated by a complete ESMFold2 forward pass, allowing the search to combine gradient-informed mutation preferences with evidence accumulated from previously verified designs.

We evaluate ProteoCraft on de novo binder design, TCR-facing binder design, and localized CDR redesign. Across all settings, ProteoCraft achieves better or competitive average design quality and the yield of high-confidence candidates over baselines and compute-matched variants of our pipeline. These improvements persist under independent AF2-IG evaluation and Rosetta interface-scoring evaluation, suggesting that the gains extend beyond the search-time verifier. Ablations demonstrate complementary contributions from truncated diffusion guidance and tree search. Finally, wet-lab validation confirms binding of selected designs to IFNAR2 and PD-L1, with BLI-measured dissociation constants of 153 nM and 0.879 nM, respectively, without experimental affinity maturation.

\begin{figure}[t]
\centering
\includegraphics[width=\linewidth]{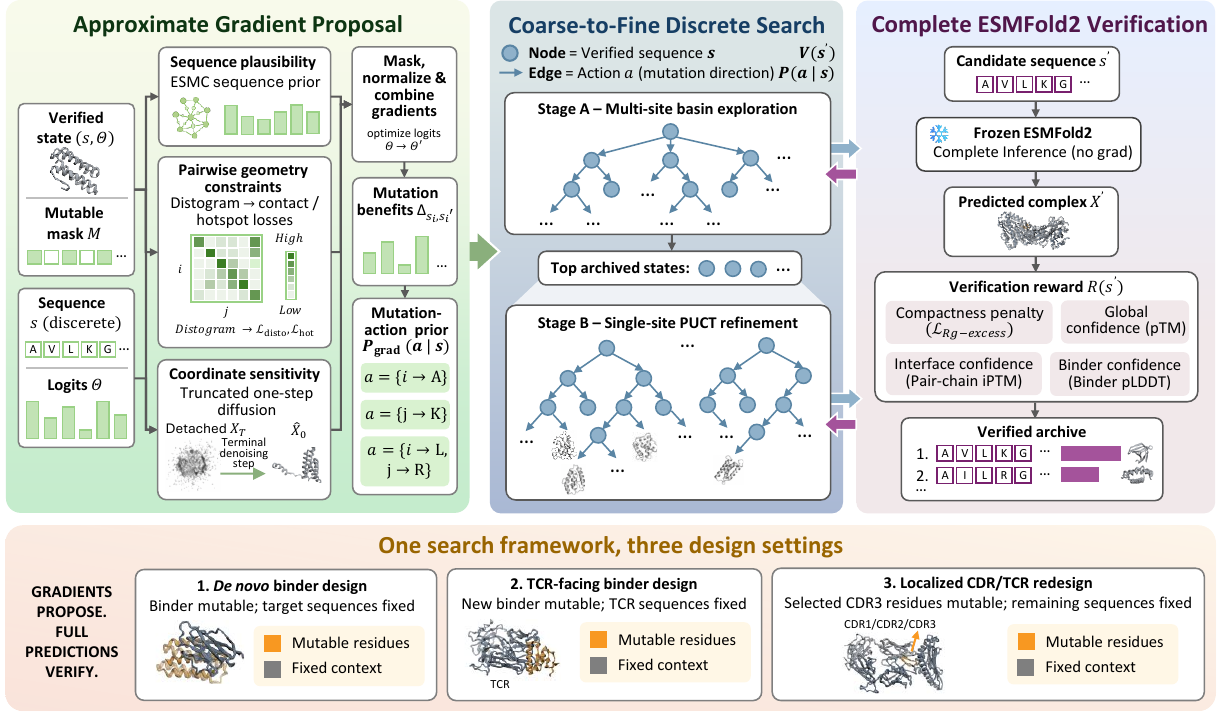}
\caption{\textbf{Overview.}
ProteoCraft formulates protein design as proposal--verification search.
Given a mutable design region and molecular context, a coarse-to-fine tree search uses truncated diffusion signals to propose candidate sequences, while complete ESMFold2 inference assigns node values.} %  Given a mutable design region and molecular context, a coarse-to-fine tree search uses truncated diffusion signals to propose candidate sequences, while complete ESMFold2 inference assigns node values. 
\label{fig}
\vspace{-10pt}
\end{figure}

% Our contributions are:
% \begin{itemize}[leftmargin=*, nosep]
% \item \textbf{Protein design as proposal--verification search.}
% We separate local differentiable guidance from complete candidate evaluation: approximate gradients define a proposal policy over discrete sequence edits, while full frozen predictor inference assigns node values.

% \item \textbf{Gradient-prior, coarse-to-fine tree search.}
% We convert complementary geometric, sequence, and truncated diffusion sensitivities into mutation priors and integrate them with verifier-backed search values. Multi-site exploration discovers promising sequence basins, followed by verified single-site refinement.

% \item \textbf{A unified algorithm across design regimes and evaluators.}
% We apply the same search formulation to de novo binder design and localized immune-receptor sequence redesign, and show improvements under both the search-time verifier and independent AF2-IG and Rosetta cross-evaluation.

% \end{itemize}

%% file: sec/3_method.tex
\section{Method}
\label{sec:method}

\subsection{Preliminary}

\paragraph{Problem Statement.} Given a target receptor $\mathcal T$, we aim to design a binder sequence $\mathbf s=(s_1,\ldots,s_L)$. A binary mask $\mathbf M\in\{0,1\}^{L}$ specifies the mutable positions, where $M_i=0$ indicates a fixed location. The binder identity is parameterized by amino-acid logits $\boldsymbol{\Theta}\in\mathbb R^{L\times 20}$, with continuous probabilities $P_{\boldsymbol{\Theta}}(i,s)=\mathrm{Softmax}(\boldsymbol{\Theta}_i/T)_s$ and discrete type $s_i=\arg\max_s\boldsymbol{\Theta}_{i,s}$, where $T>0$ is an annealed temperature and $P_{\boldsymbol{\Theta}}$ exposes differentiable sequence conditioning. 

\paragraph{Tree-based Search for Protein Discovery.} ProteoCraft adopts a modified MCTS framework~\citep{wu2026deepsearch} to build a search tree for \emph{de novo} binder design. It regards the protein design process as a finite-horizon Markov Decision Process (MDP), represented by the tuple $\langle S, A, P, R, \gamma \rangle$. Specifically, $S$ is the set of states that contains the entire binder space. $A$ represents possible actions, where each action $a$ corresponds to a different mutation direction. $P(\mathbf{s}' \mid \mathbf{s}, a)$ denotes the transition probability of reaching state $\mathbf{s}'$ from $\mathbf{s}$ after action $a$, $R(\mathbf{s})$ is the reward function to evaluate the quality of the new mutant sequences $\mathbf{s}$, and we use undiscounted reward backup, corresponding to $\gamma=1$. The root node $\mathbf{s}^0$ represents the wild-type or random binder sequence, and each child node $\mathbf{s}^i$ corresponds to a mutant variant, for which we assign a q-value $q(\mathbf{s}^i)$. Then, a root-to-leaf path ending at a terminal node $\mathbf{s}^\text{end}$ forms a trajectory $\mathbf{t} = \mathbf{s}^0 \oplus\mathbf{s}^1 \oplus\mathbf{s}^2 \oplus \ldots \oplus\mathbf{s}^\text{end}$, which displays an artificial evolutionary pathway from $\mathbf{s}^0$. 

% separates protein design into two complementary operations: \emph{proposal} and \emph{verification}. Differentiable signals from a diffusion-based structure predictor and a protein language model determine promising sequence perturbations and propose discrete edits. Candidate quality is then scored by complete inference of the structure predictor. A tree search connects these two operations and allocates expensive verification calls across sequence space. 

\subsection{Expansion: From Structural Gradients to Mutation Proposals}
\label{sec:gradient_proposals}

ProteoCraft iteration consists of four phases: \emph{selection}, \emph{expansion}, \emph{simulation}, and \emph{backpropagation}. A search node stores a verified discrete sequence together with its inherited logits, predicted structure, verification metrics, and search statistics. For a selected verified node $\mathbf{s}^i$, proposal optimization starts from its inherited logits. We pass the corresponding probability matrix $P_{\boldsymbol{\Theta}}$ through the ESMFold2 \texttt{res\_type\_soft} interface and combine gradients for pairwise geometry, sequence plausibility, and coordinate-level structural sensitivity. Only mutable positions receive logit updates.

We consider three categories of mutation optimization objectives. The foremost type is the folding model distogram prior, formed as $\mathcal L_{\mathrm{disto}}(\mathbf{s}^i)=0.5\mathcal L_{\mathrm{intra}}+0.5\mathcal L_{\mathrm{inter}}+0.2\mathcal L_{\mathrm{glob}}$, where $\mathcal L_{\mathrm{intra}}$ promotes coherent binder-internal contacts, $\mathcal L_{\mathrm{inter}}$ promotes target--binder contacts, and $\mathcal L_{\mathrm{glob}}$ regularizes global compactness. Secondly, we use an ESMC masked language-model loss $\mathcal L_{\mathrm{ESMC}}$~\citep{candido2026language}. We mask sampled mutable positions and compute cross-entropy against the current soft amino-acid probabilities at those positions. This loss regularizes sequence plausibility during proposal optimization.

Though distogram gradients from $\mathcal L_{\mathrm{disto}}(\mathbf{s}^i)$ capture pairwise geometric preferences, they fail to directly expose sensitivity through the coordinate-generating process. To bridge this gap, we differentiate through a single terminal diffusion step to acquire the extra coordinate-sensitive mutation signal. Let $\mathbf{X}_p$ denote the atom coordinates of the modeled complex at noise level $p$, and let $f_\phi^{p\rightarrow q}$ denote the reverse-diffusion trajectory from noise level $p$ to $q$. The model parameters $\phi$ remain frozen. Starting from $\mathbf{X}_{\sigma_{\max}}=\sigma_{\max}\boldsymbol{\epsilon}$ with $\boldsymbol{\epsilon}\sim\mathcal N(\mathbf{0},\mathbf{I})$, we run the trajectory without gradient tracking until noise level $\tau$, then apply one differentiable denoising operation:
\begin{equation}
\begin{aligned}
    \mathbf{X}_\tau
    &=\operatorname{sg}\!\left[f_\phi^{\sigma_{\max}\rightarrow\tau}
    (\mathbf{X}_{\sigma_{\max}};\mathbf{c}_{\boldsymbol{\Theta}})\right],\\
    \hat{\mathbf{X}}_0
    &=D_\phi(\mathbf{X}_\tau,\tau;\mathbf{c}_{\boldsymbol{\Theta}}).
\end{aligned}
\label{eq:one_step}
\end{equation}
Here, $\mathbf{c}_{\boldsymbol{\Theta}}$ is the sequence-conditioned representation, $\operatorname{sg}$ denotes stop-gradient, and $D_\phi$ predicts clean coordinates at noise level $\tau$. Gradients are retained through the terminal denoising operation and the current sequence conditioning, while the preceding reverse-diffusion trajectory is detached. Here, we define $\mathcal L_{\mathrm{1\, step}}\left(\hat {\mathbf{X}}_0\right)= 0.1\mathcal L_{\mathrm{pLDDT}}+0.4\mathcal L_{\mathrm{PAE}} +0.2\mathcal L_{\mathrm{iPAE}} +0.3\mathcal L_{\mathrm{pair\text{-}chain\,iPTM}} +0.1\mathcal L_{\mathrm{Rg}} +0.1\mathcal L_{\mathrm{clash}} +0.1\mathcal L_{\mathrm{i\text{-}clash}}$, where lower values are preferred for all terms. 
For $\mu\in\{\mathrm{disto},\mathrm{ESMC},\mathrm{1\,step}\}$, let $\mathbf{g}_\mu$ denote $\nabla_{\boldsymbol{\Theta}}\mathcal L_\mu$ after zeroing the rows corresponding to fixed positions. We normalize each masked gradient separately and obtain the combined proposal gradient $\mathbf{g}=\operatorname{Norm}(\mathbf{g}_{\mathrm{disto}})+\lambda_{\mathrm{ESMC}}\operatorname{Norm}(\mathbf{g}_{\mathrm{ESMC}})+\lambda_{\mathrm{1\,step}}\operatorname{Norm}(\mathbf{g}_{\mathrm{1\,step}})$. The normalization and update schedule are specified in App.~\ref{app:implementation}.

Starting from the inherited logits, we perform multiple updates of the form $\boldsymbol{\Theta}\leftarrow\boldsymbol{\Theta}-\eta\mathbf{g}$. Let $\boldsymbol{\Theta}$ and $\boldsymbol{\Theta}'$ denote the logits before and after this proposal-optimization phase, respectively. In Stage A, the raw argmax sequence $\mathbf{s}'=\arg\max(\boldsymbol{\Theta}')$ is retained as a candidate and must undergo complete verification before being considered as a final design. We also use the changes in logits to construct discrete mutation proposals. For a substitution $s_i\rightarrow s_i'$ at position $i$, we define
\begin{equation}
    \boldsymbol{\Delta}_{s_i,s_i'}= \left(\boldsymbol{\Theta}'_{i,s_i'}-\boldsymbol{\Theta}_{i,s_i'}\right) -\left(\boldsymbol{\Theta}'_{i,s_i}-\boldsymbol{\Theta}_{i,s_i}\right).
\label{eq:mutation_benefit}
\end{equation}
A positive $\Delta_{s_i,s_i'}$ indicates that the combined hallucination update increasingly favors residue $s_i'$ over retaining $s_i$. For an admissible single-site action $a=(i,s_i')$, where $M_i=1$ and $s_i'\neq s_i$, we write $\Delta_a=\Delta_{s_i,s_i'}$. The gradient-derived proposal prior is
\[
P_{\mathrm{grad}}(a\mid\mathbf{s})
=
\frac{\exp(\Delta_a)}
{\sum_{a'\in\mathcal A(\mathbf{s})}\exp(\Delta_{a'})},
\]
where $\mathcal A(\mathbf{s})$ contains the admissible single-site mutation actions from $\mathbf{s}$. Multi-site candidates are constructed from mutation benefits as described in Section~\ref{sec:tree_search}. Applying a proposed mutation action $a$ to $\mathbf{s}$ produces a discrete candidate $\mathbf{s}'$. Candidates undergo complete ESMFold2 verification before selected candidates are inserted as children of $\mathbf{s}$. Each inserted child inherits the optimized logits $\boldsymbol{\Theta}'$ from the proposal that generated it, rather than the logits before optimization. When necessary, these inherited logits are minimally aligned so that their argmax matches the verified child sequence.

\subsection{Evaluation and Backpropagation with Full-model Verifier}
\label{sec:verification}

Each discrete candidate is evaluated by complete frozen ESMFold2 inference without gradient tracking. The baseline verifier loss is $\mathcal L_{\mathrm{verify}}(\mathbf{s})=(1-\mathrm{pair\text{-}chain\ iPTM})+0.2(1-\mathrm{binder\ pLDDT})+0.1(1-\mathrm{pTM})+0.1\mathcal L_{\mathrm{Rg\text{-}excess}}$, where lower values are preferred. Pair-chain iPTM is computed between the designated target chains and the binder, while $\mathcal L_{\mathrm{Rg\text{-}excess}}$ penalizes excessive binder radius of gyration relative to a length-dependent threshold. Additional structural and hotspot statistics are recorded separately. In Stage A, the verification reward is $R(\mathbf{s})=-\mathcal L_{\mathrm{verify}}(\mathbf{s})$.

All successful full-model evaluations are retained in an archive, while only selected candidates are inserted into the tree. Rewards from these selected candidates are backed up along the selected search path without depth discounting. The corresponding visit counts and empirical mean values are updated to guide which sequence branches receive further exploration. These updates do not propagate verifier gradients into proposal optimization. Implementation details are in App.~\ref{app:implementation}.

\subsection{Coarse-to-Fine Multi-stage Selection Strategy}
\label{sec:tree_search}

Starting from the root $\mathbf{s}^0$, we progressively adapt the granularity of sequence editing through two regimes: \emph{multi-site exploration} and \emph{single-site refinement}. Early in search, coordinated multi-site mutations provide a larger effective search radius, enabling efficient transitions across distinct sequence basins. After the allocated Stage-A budget is exhausted, the search switches to single-site substitutions for local refinement around selected verified sequences.

\textbf{Stage A: multi-site basin exploration} begins from multiple independently initialized sequence--logit states. At each iteration, we first allocate search budget across the initialized roots and then select a verified node within the chosen tree using the Upper Confidence Bound (UCB) criterion~\citep{kocsis2006bandit}:
\begin{equation}
    U(\mathbf{s}^i)=q(\mathbf{s}^i)+c_{\mathrm{ucb}}\sqrt{\frac{\log N_{\mathrm{total}}}{N(\mathbf{s}^i)+\epsilon}},
    \label{eq:ucb}
\end{equation}
where $q(\mathbf{s}^i)$ is the mean verifier reward backed up through node $\mathbf{s}^i$, $N(\mathbf{s}^i)$ is its visit count, and $N_{\mathrm{total}}$ is the total number of visits. Since rewards are defined as negative verifier losses in Stage A, larger $q(\mathbf{s}^i)$ indicates better verified outcomes. The UCB score balances exploitation of promising states with exploration of less-visited regions.
After selecting a node, we optimize its inherited logits and construct multi-site candidates from the mutation benefits in Eq.~\ref{eq:mutation_benefit}. For these benefit-based candidates, the maximum number of edited residues is $K_{\max}=\lceil\rho\sum_i M_i\rceil$. The raw argmax proposal is retained separately and is not constrained by $K_{\max}$. Each candidate undergoes complete verification before selected candidates are inserted into the tree.

\textbf{Stage B: single-site refinement.} After Stage A, selected verified sequence--logit states are re-evaluated with a common verifier seed to initialize Stage-B roots. Each newly proposed candidate differs from its parent at a single residue. Stage B uses a balanced reward derived from full-verifier metrics, with additional penalties for unmet hotspot constraints, when specified, and interfaces below a quality floor. This reward guides which sequence branches are explored further. Let $P(a\mid\mathbf{s}^i)$ denote the action prior used by Stage-B search, $q(\mathbf{s}^i,a)$ the mean Stage-B reward backed up through action $a$, and $N(\mathbf{s}^i,a)$ the corresponding backup count. PUCT guides traversal of existing mutation branches~\citep{rosin2011multi}:
\begin{equation}
    U_{\mathrm{PUCT}}(\mathbf{s}^i,a)
    =
    q(\mathbf{s}^i,a)
    +c_{\mathrm{puct}}P(a\mid\mathbf{s}^i)
    \frac{\sqrt{N(\mathbf{s}^i)}}{1+N(\mathbf{s}^i,a)}.
    \label{eq:puct}
\end{equation}
At the selected expansion state, we optimize the inherited logits and construct single-site mutation candidates. These candidates undergo complete verification before selected candidates are added to the tree. Their rewards guide subsequent selection between mutation branches. Thus, single-site refinement restricts the number of changed residues per candidate, not the number of candidates evaluated per iteration.

\subsection{Constrained and Localized Design}
\label{sec:localized_design}

ProteoCraft supports different design settings by modifying $\mathbf{M}$ and the interface terms in the optimization objective $\mathcal L_{\mathrm{disto}}(\mathbf{s})$. For \emph{de novo} and TCR-facing binder design, all binder residues are mutable while target sequences remain fixed and provide structural context. For localized CDR redesign, the remaining framework sequence participates in structure prediction but is not updated. The framework sequence remains fixed, but its coordinates are not explicitly clamped. % Because ESMFold2 does not consume an external template structure, localized redesign preserves the framework sequence rather than explicitly clamping its input backbone coordinates.

When target hotspots $\mathcal H\subseteq\mathcal T$ are specified, we define the hotspot constraint loss as $\mathcal L_{\mathrm{hot}}= \frac{1}{|\mathcal H|}\sum_{i\in \mathcal H}\min_{j\in \mathcal B}\ell^{\mathrm{contact}}_{ij}(d_{\mathrm{hot}})$, where $\mathcal B$ denotes the set of binder residue indices, $\ell^{\mathrm{contact}}_{ij}$ is the pairwise contact loss, and $d_{\mathrm{hot}}$ is the predefined contact cutoff distance. Then, we replace the standard inter-chain loss term with $\mathcal L_{\mathrm{inter}}'=(1-\lambda_{\mathrm{hot}})\mathcal L_{\mathrm{inter}}+\lambda_{\mathrm{hot}}\mathcal L_{\mathrm{hot}}$ to encourage the interaction between hotspot residues and the candidate binder. Via this mechanism, hotspot information therefore biases the mutation direction toward a desired target region $\mathcal H$. 
% The full verifier additionally records hotspot distance and coverage, which may be used for final candidate selection without entering the default verifier objective.

%% file: sec/4_experiment.tex
\section{Experiments}
\label{sec:experiments}

% We evaluate ProteoCraft around three questions:
% (i) whether gradient-informed proposal--verification search improves protein design under a matched candidate budget;
% (ii) whether the improvement persists under evaluators that are independent of the search-time verifier; and
% (iii) which components of the proposal and search procedure are responsible for the gain.
% We first study \emph{de novo} binder design at scale, then evaluate immune-relevant TCR-facing binder design and localized CDR/TCR redesign, and finally analyze the contribution and robustness of the proposal mechanism.

\subsection{Experimental Setup}
\label{subsec:exp_setup}

\paragraph{Design tasks.}
We consider three design settings. (1) \emph{De novo} binder design, where the complete binder sequence is generated against a fixed target. Our main benchmark comprises 14 target--hotspot settings across eight target proteins, spanning cellular receptors, allergens, and other structurally diverse proteins.
For each method, we generate 15 candidates per setting using three binder lengths and five random seeds, yielding $N=210$ candidates per method. (2) Immune-relevant TCR-facing binder design, where a new binder is designed against a fixed TCR target. (3) Localized CDR/TCR redesign, where only a designated CDR region is mutable while the remaining receptor sequence provides fixed structural context.
Detailed targets, sequence lengths, hotspot annotations, and generation settings are provided in Appx.~\ref{app:experimental_details}.

\paragraph{Baselines.}
We compare against multiple state-of-the-art algorithms, containing PXDesign-d~\citep{team2025pxdesign}, RFdiffusion v3~\citep{butcher2025novo}, Proteina-Complexa~\citep{didi2026scaling}, BoltzGen~\citep{stark2025boltzgen}, BindCraft~\citep{pacesa2025one}, BindCraft v2~\citep{bindcraft2}, and BoltzDesign1~\citep{cho2025boltzdesign1}. We further formulate three matched variants of our own pipeline:
\emph{No One-step Diffusion} removes the truncated One-step diffusion gradient while retaining the remaining proposal--verification procedure; \emph{ProteoCraft-Best-of-$N$} uses the same differentiable proposal mechanism and matched candidate budget but selects among independent proposals without tree-search control; and \emph{ProteoCraft-MCTS} uses the full coarse-to-fine tree search. These comparisons separate the contribution of diffusion-based proposal guidance from the contribution of verifier-guided search. All comparisons use matched target inputs and evaluation protocols whenever supported by the corresponding baseline. Notably, for BindCraft v2 on TCR-related binder design, no candidates passed its built-in filters within our sampling budget.

\paragraph{Evaluation.} We propose \emph{model-aligned} evaluation and \emph{independent cross-evaluation}.
First, complete frozen ESMFold2 inference reports Pair-chain iPTM, complex iPTM, pTM, pLDDT, and binder-specific confidence.
Second, we re-evaluate the same discrete sequences using AF2-IG~\citep{watson2023novo} as an independent structure verifier.
Third, we use a common Rosetta~\citep{alford2017rosetta} interface-scoring pipeline to measure interface energetics, shape complementarity, packing, hydrogen bonding, and filter pass rates.
This separation is important because the search-time verifier is used to allocate search, whereas AF2-IG and Rosetta test whether improvements persist outside the optimization model.
Full evaluation definitions and thresholds are in Appx.~\ref{app:evaluation_protocol}.
\begin{table*}[t]
\centering
\scriptsize
\caption{\textbf{ESMFold2 evaluation of \emph{de novo} binder design on 14 target--hotspot settings across eight target proteins.}
Each method contributes 15 candidates per hotspot ($N=210$).
The left block reports mean confidence metrics (higher is better);
the right block reports high-confidence design yields as count (percentage).
The strict criterion requires both Pair iPTM $\geq 0.75$
and Binder pLDDT $\geq 0.9$.
Dashes indicate results not yet reported.}
\vspace{-3pt}
\label{tab:denovo_esmf2}
\label{tab:denovo_threshold}
\setlength{\tabcolsep}{4pt}
\renewcommand{\arraystretch}{1.12}
\resizebox{\textwidth}{!}{%
\begin{tabular}{lccccc ccc}
\toprule
Method & \multicolumn{5}{c}{\textbf{Mean confidence} $\uparrow$}
& \multicolumn{3}{c}{\textbf{High-confidence yield} $\uparrow$} \\
\cmidrule(lr){2-6}
\cmidrule(lr){7-9}
& Pair iPTM & iPTM & pTM & pLDDT & Binder pLDDT
& \shortstack{Pair iPTM\\$\geq 0.5$}
& \shortstack{Pair iPTM\\$\geq 0.7$}
& \shortstack{Pair iPTM $\geq 0.75$\\Binder pLDDT $\geq 0.9$} \\
\midrule
RFdiffusion v3
& 0.206 & 0.282 & 0.593 & 0.739 & 0.729
& 20 (9.5\%) & 1 (0.5\%) & 0 (0.0\%) \\
BoltzGen
& 0.438 & 0.577 & 0.748 & 0.826 & 0.857
& 105 (50.0\%) & 45 (21.4\%) & 9 (4.3\%) \\
Proteina-Complexa & 0.286 & 0.386 & 0.653 & 0.774 & 0.779 & 48 (22.9\%) & 4 (1.9\%) & 0 (0.0\%) \\ 
PXDesign-d
& 0.322 & 0.433 & 0.683 & 0.801 & 0.832
& 58 (27.6\%) & 16 (7.6\%) & 1 (0.5\%) \\
BoltzDesign1 & 0.154 & 0.220 & 0.562 & 0.638 & 0.490 & 6 (2.9\%) & 0 (0.0\%) & 0 (0.0\%) \\ 
BindCraft
& \underline{0.573}
& \underline{0.779}
& \underline{0.822}
& \underline{0.833}
& \underline{0.858}
& \underline{162} (77.1\%)
& \underline{53} (25.2\%)
& \underline{11} (5.2\%) \\
BindCraft v2 & 0.347 & 0.470 & 0.691 & 0.766 & 0.740 & 70 (33.3\%) & 7 (3.3\%) & 1 (0.5\%) \\
\midrule
No One-step Diffusion
& 0.637 & 0.861 & 0.869 & 0.859 & 0.879
& 176 (83.8\%) & 85 (40.5\%) & 38 (18.1\%) \\
ProteoCraft-Best-of-$N$
& 0.660 & 0.885 & 0.891 & 0.866 & 0.880
& 186 (88.6\%) & 118 (56.2\%) & 40 (19.0\%) \\

ProteoCraft-Stage-A-only & 0.691 & 0.906 & 0.895 & 0.876 & 0.906 & {195 (92.9\%)} & 138 (65.7\%) & 85 (40.5\%) \\
\textbf{ProteoCraft-MCTS} & \textbf{0.721} & \textbf{0.930} & \textbf{0.913} & \textbf{0.889} & \textbf{0.917} & \textbf{195 (92.9\%)} & \textbf{165 (78.6\%)} & \textbf{114 (54.3\%)} \\
\bottomrule
\end{tabular}%
}
\vspace{-10pt}
\end{table*}

\subsection{\emph{De Novo} Binder Design}
\label{subsec:main_results}

\paragraph{Model-aligned Evaluation.} Tab.~\ref{tab:denovo_esmf2} reports ESMFold2 evaluation over all 210 candidates, where ProteoCraft-MCTS obtains the strongest result across each metric.
Relative to the compute-matched Best-of-$N$ baseline, tree-search control improves mean Pair-chain iPTM from $0.660$ to $0.721$ and Binder pLDDT from $0.880$ to $0.917$.
Removing the One-step diffusion proposal signal further degrades Pair-chain iPTM to $0.637$, suggesting that the coordinate-level diffusion gradient contributes information beyond the remaining geometric and sequence proposal terms.

Notice that mean scores can hide whether a method consistently produces usable designs. We therefore additionally measure the fraction of candidates satisfying increasingly selective interface-confidence criteria in Tab.~\ref{tab:denovo_threshold}. The gain becomes larger as the criterion becomes more selective. Under the strictest threshold, ProteoCraft-MCTS yields $114/210$ high-quality candidates ($54.3\%$), compared with $40/210$ ($19.0\%$) for the compute-matched Best-of-$N$ baseline. Thus, tree search does more than improve the average verifier score: it substantially increases the probability that a finite design budget produces a candidate satisfying both interface and folding-confidence requirements.

Per-target results are provided in Appx.~\ref{tab:per_target_pairiptm}.
ProteoCraft-MCTS improves Pair-chain iPTM over external baselines on essentially all targets and achieves the strongest mean result among the matched ProteoCraft variants on most targets. The remaining difficult cases also illustrate why complete verification is retained throughout search: \emph{proposal guidance improves where the algorithm searches, but individual targets can still exhibit predictor-specific uncertainty or objective mismatch}.

\paragraph{Independent AF2-IG cross-evaluation.}
Tab.~\ref{tab:denovo_af2ig_rosetta} evaluates the same $210$ sequences from each method using AF2-IG~\citep{watson2023novo}. ProteoCraft-MCTS again obtains the strongest interface metrics and the highest strict pass rate. The improvement therefore persists under a different structure-prediction model: the strict AF2-IG pass rate increases from $78.6\%$ for Best-of-$N$ to $86.2\%$ for ProteoCraft-MCTS. This result argues against interpreting the ESMFold2 gains solely as optimization of the internal verifier.

\paragraph{Rosetta interface evaluation.}
We next test with a common Rosetta pipeline, whose scores assess independently refined interface energetics and geometry. Tab.~\ref{tab:denovo_af2ig_rosetta} shows that ProteoCraft-MCTS produces the most favorable mean interface energy, energy normalized by buried surface area, and hydrogen-bond count, while matching the highest shape complementarity. Although BoltzGen obtains the highest mean Packstat, the overall pattern is consistent with the AF2-IG cross-evaluation. Improvements obtained through proposal--verification search persist under evaluators that are not used to guide the search.
% The corresponding Low/Medium/High Rosetta filter yields are reported in Appx.~\ref{tab:app_rosetta_filters}. The High-filter pass rate is $30.0\%$ for ProteoCraft-MCTS, compared with $21.9\%$ for Best-of-$N$ and $31.4\%$ for the No-One-step variant.
% This difference also emphasizes that no single external filter is perfectly aligned with the ESMFold2 search objective.

\begin{table*}[t]
\centering
\scriptsize
\caption{\textbf{Independent AF2-IG and Rosetta evaluation of \emph{de novo} designs.} Each was evaluated on $N=210$ designs. For AF2-IG, lower iPAE and higher iPTM, pTM, and strict pass rate are better.
For Rosetta, lower $\Delta G$ and $\Delta G/\Delta \mathrm{SASA}$ and higher values for the remaining metrics are better.
Bold and underlined values indicate the best and second-best results, respectively.}
\label{tab:denovo_af2ig_rosetta}
\resizebox{\textwidth}{!}{%
\begin{tabular}{l|cccc|cccccc}
\toprule
& \multicolumn{4}{c}{AF2-IG} & \multicolumn{6}{c}{Rosetta interface evaluation} \\
\cmidrule(lr){2-5} \cmidrule(lr){6-11}
Method & iPAE $\downarrow$ & iPTM $\uparrow$ & pTM $\uparrow$ & Strict Pass $\uparrow$ & $\Delta G \downarrow$ & $\Delta \mathrm{SASA} \uparrow$ & SC $\uparrow$ & $\Delta G/\Delta \mathrm{SASA} \downarrow$ & H-bonds $\uparrow$ & Packstat $\uparrow$ \\
\midrule
RFdiffusion v3 & 25.426 & 0.142 & 0.550 & 0 (0.0\%) & -49.398 & 1834.2 & 0.633 & -2.696 & 4.662 & 0.612 \\
BoltzGen & 26.342 & 0.200 & 0.689 & 0 (0.0\%) & -31.138 & 1902.4 & 0.558 & -1.640 & 6.463 & \textbf{0.654} \\
Proteina-Complexa & 18.927 & 0.359 & 0.651 & 6 (2.9\%) & -28.741 & 1626.0 & 0.557 & -1.671 & 5.314 & 0.597 \\
PXDesign-d & 18.949 & 0.349 & 0.650 & 20 (9.5\%) & -50.477 & 2029.6 & 0.603 & -2.469 & \underline{8.595} & 0.597 \\
BoltzDesign1 & 19.967 & 0.369 & 0.571 & 1 (0.5\%) & -52.903 & \textbf{3251.0} & 0.607 & -1.611 & 7.743 & 0.542 \\
BindCraft & 7.093 & 0.807 & \underline{0.841} & 104 (49.5\%) & \underline{-57.879} & 2085.9 & \underline{0.665} & \underline{-2.778} & 8.500 & 0.613 \\
BindCraft v2 & \underline{7.062} & \underline{0.804} & 0.835 & \underline{122} (58.1\%) & -49.981 & 2029.1 & 0.644 & -2.469 & 6.933 & \underline{0.621} \\
\midrule
No One-step Diffusion & 6.162 & 0.848 & 0.864 & 126 (60.0\%) & {-70.665} & 2585.6 & \textbf{0.681} & -2.760 & 11.095 & 0.583 \\
ProteoCraft-Best-of-$N$ & {5.795} & {0.860} & {0.881} & {165 (78.6\%)} & -68.078 & 2666.7 & 0.641 & -2.588 & {11.395} & 0.576 \\
\textbf{ProteoCraft-MCTS} & \textbf{4.792} & \textbf{0.883} & \textbf{0.892} & \textbf{181 (86.2\%)} & \textbf{-81.195} & \underline{2933.7} & \textbf{0.681} & \textbf{-2.783} & \textbf{12.510} & 0.584 \\
\bottomrule
\end{tabular}%
}
\end{table*}

\subsection{Immune-Relevant Protein Discovery}
\label{subsec:tcr_binder}

\paragraph{TCR-Facing Binder Design.} We evaluate ProteoCraft under TCR-facing surfaces, which differ geometrically from the standard \emph{de novo} benchmark. Tab.~\ref{tab:tcr_binder} reports matched ESMFold2 and AF2-IG evaluation. ProteoCraft-MCTS improves over the compute-matched Best-of-$N$ baseline on all reported ESMFold2 metrics, including Pair-chain iPTM from $0.514$ to $0.634$ and pLDDT from $0.866$ to $0.889$. The direction of improvement is also preserved under AF2-IG, where MCTS obtains lower iPAE and higher iPTM and pTM.
These results suggest that proposal--verification search is not restricted to conventional globular binder targets.

\begin{table}[t]
\centering
\small
\caption{\textbf{TCR-facing binder design.} The left block reports complete ESMFold2 verification and the right block reports independent AF2-IG evaluation.}
\label{tab:tcr_binder}
\resizebox{0.8\linewidth}{!}{
\begin{tabular}{l|cccc|ccc}
\toprule
& \multicolumn{4}{c|}{ESMFold2}
& \multicolumn{3}{c}{AF2-IG} \\
Method
& Pair iPTM $\uparrow$ & iPTM $\uparrow$ & pTM $\uparrow$ & pLDDT $\uparrow$
& iPAE $\downarrow$ & iPTM $\uparrow$ & pTM $\uparrow$ \\
\midrule
RFdiffusion v3 & 0.225 & 0.695 & 0.780 & 0.852 & 27.540 & 0.161 & 0.693 \\
BoltzGen & 0.384 & 0.763 & 0.824 & 0.843 & 26.342 & 0.200 & 0.689 \\
Proteina-Complexa & {0.300} & {0.756} & {0.822} & {0.863} & {26.439} & {0.191} & {0.702} \\
PXDesign-d & 0.276 & 0.717 & 0.795 & 0.873
& \underline{26.115} & \underline{0.207} & 0.708 \\
BoltzDesign1 & {0.228} & {0.734} & {0.810} & {0.817} & {26.165} & {0.214} & {0.713} \\
BindCraft
& \underline{0.513} & \underline{0.828} & \underline{0.869} & \underline{0.884}
& 27.043 & 0.172 & \underline{0.718} \\
% BindCraft v2 & \textcolor{blue}{0.320} & \textcolor{blue}{0.766} & \textcolor{blue}{0.829} & \textcolor{blue}{0.856} & \textcolor{blue}{24.996} & \textcolor{blue}{0.249} & \textcolor{blue}{0.729} \\
\midrule
ProteoCraft-Best-of-$N$
& 0.514 & 0.841 & 0.877 & 0.866
& 25.266 & 0.249 & 0.725 \\
\textbf{ProteoCraft-MCTS}
& \textbf{0.634} & \textbf{0.887} & \textbf{0.907} & \textbf{0.889}
& \textbf{24.595} & \textbf{0.282} & \textbf{0.730} \\
\bottomrule
\end{tabular}}
\end{table}

% TODO: insert the updated TCR Rosetta cross-evaluation here once finalized.

\paragraph{Localized CDR/TCR Redesign.}
\label{subsec:cdr_only}

ProteoCraft can also operate on a restricted mutable region while retaining the rest of the complex as structural context.
We evaluate four chain-specific redesign settings from two TCR complexes, with ten candidates per setting. 
Tab.~\ref{tab:cdr_esmf2} demonstrates that ProteoCraft-MCTS obtains the strongest mean peptide-specific and context-averaged interface confidence.
Relative to RFdiffusion v3, mean Peptide Pair iPTM increases from $0.535$ to $0.765$. The improvement is particularly pronounced on the harder 1QRN $\beta$ setting, where ProteoCraft raises Peptide Pair iPTM from $0.421$ to $0.789$.

We again cross-evaluate these designs using AF2-IG and Rosetta in Tab.~\ref{tab:cdr_cross_eval}. ProteoCraft-MCTS achieves the lowest AF2-IG iPAE and highest iPTM, while Rosetta reports substantially more favorable mean interface energy, shape complementarity, hydrogen bonding, and packing. Together, these results indicate that restricting the mutable sequence region does not prevent proposal--verification search from exploiting global structural context. 

\begin{table}[t]
\centering
\scriptsize
\caption{\textbf{ESMFold2 evaluation for localized CDR redesign.} Peptide Pair iPTM measures confidence between the redesigned chain and the peptide; Mean Pair iPTM averages pairwise confidence between the redesigned chain and all fixed context chains.
All metrics are higher-is-better.}
\label{tab:cdr_esmf2}
\resizebox{\linewidth}{!}{%
\begin{tabular}{llccccccc}
\toprule
Method & Case & $N$ & Peptide Pair $\uparrow$
& Mean Pair $\uparrow$ & iPTM $\uparrow$ & pTM $\uparrow$
& Complex pLDDT $\uparrow$ & Chain pLDDT $\uparrow$ \\
\midrule
\multirow{5}{*}{RFdiffusion v3}
& 1QRN $\alpha$ & 10 & 0.591 & 0.686 & 0.872 & 0.896 & 0.919 & 0.901 \\
& 1QRN $\beta$  & 10 & 0.421 & 0.560 & 0.824 & 0.857 & 0.905 & 0.912 \\
& 9NMW $\alpha$ & 10 & 0.549 & 0.669 & 0.884 & 0.900 & 0.906 & 0.896 \\
& 9NMW $\beta$  & 10 & 0.579 & 0.687 & 0.909 & 0.920 & 0.914 & 0.918 \\ \cmidrule(lr){2-9}
& \textbf{Mean} & -- & 0.535 & 0.650 & 0.872 & 0.893 & 0.911 & 0.907 \\
\midrule
\multirow{5}{*}{BoltzGen}
& 1QRN $\alpha$ & 10 & 0.715 & 0.761 & 0.926 & 0.934 & 0.929 & 0.912 \\
& 1QRN $\beta$  & 10 & 0.491 & 0.609 & 0.851 & 0.875 & 0.909 & 0.917 \\
& 9NMW $\alpha$ & 10 & 0.567 & 0.675 & 0.880 & 0.898 & 0.910 & 0.897 \\
& 9NMW $\beta$  & 10 & 0.650 & 0.704 & 0.894 & 0.908 & 0.914 & 0.912 \\
\cmidrule(lr){2-9}
& \textbf{Mean} & -- & 0.606 & 0.687 & 0.888 & 0.904 & 0.915 & 0.910 \\ \midrule 
\multirow{5}{*}{BindCraft}
& 1QRN $\alpha$ & 10 & 0.752 & 0.774 & 0.935 & 0.939 & 0.922 & 0.885 \\
& 1QRN $\beta$  & 10 & 0.464 & 0.594 & 0.842 & 0.871 & 0.898 & 0.887 \\
& 9NMW $\alpha$ & 10 & 0.603 & 0.691 & 0.883 & 0.896 & 0.901 & 0.879 \\
& 9NMW $\beta$  & 10 & 0.720 & 0.754 & 0.935 & 0.939 & 0.925 & 0.928 \\\cmidrule(lr){2-9}
& \textbf{Mean} & -- & 0.635 & 0.704 & 0.899 & 0.911 & 0.911 & 0.895 \\\midrule
% \multirow{5}{*}{BindCraft v2}
% & 1QRN $\alpha$ & \textcolor{blue}{5} & \textcolor{blue}{0.672} & \textcolor{blue}{0.715} & \textcolor{blue}{0.879} & \textcolor{blue}{0.900} & \textcolor{blue}{0.921} & \textcolor{blue}{0.904} \\
% & 1QRN $\beta$ & \textcolor{blue}{5} & \textcolor{blue}{0.546} & \textcolor{blue}{0.631} & \textcolor{blue}{0.847} & \textcolor{blue}{0.872} & \textcolor{blue}{0.908} & \textcolor{blue}{0.912} \\
% & 9NMW $\alpha$ & \textcolor{blue}{5} & \textcolor{blue}{0.618} & \textcolor{blue}{0.677} & \textcolor{blue}{0.858} & \textcolor{blue}{0.877} & \textcolor{blue}{0.902} & \textcolor{blue}{0.873} \\
% & 9NMW $\beta$ & \textcolor{blue}{5} & \textcolor{blue}{0.669} & \textcolor{blue}{0.696} & \textcolor{blue}{0.887} & \textcolor{blue}{0.903} & \textcolor{blue}{0.913} & \textcolor{blue}{0.911} \\
% \cmidrule(lr){2-9}
% & \textbf{Mean} & -- & \textcolor{blue}{0.626} & \textcolor{blue}{0.680} & \textcolor{blue}{0.868} & \textcolor{blue}{0.888} & \textcolor{blue}{0.911} & \textcolor{blue}{0.900} \\ \midrule

\multirow{5}{*}{\textbf{ProteoCraft-MCTS}}
& 1QRN $\alpha$ & 10 & 0.816 & 0.818 & 0.951 & 0.953 & 0.939 & 0.919 \\
& 1QRN $\beta$  & 10 & 0.789 & 0.779 & 0.938 & 0.943 & 0.932 & 0.942 \\
& 9NMW $\alpha$ & 10 & 0.682 & 0.732 & 0.922 & 0.923 & 0.913 & 0.880 \\
& 9NMW $\beta$  & 10 & 0.772 & 0.777 & 0.939 & 0.943 & 0.928 & 0.930 \\\cmidrule(lr){2-9}
& \textbf{Mean} & --
& \textbf{0.765} & \textbf{0.776}
& \textbf{0.937} & \textbf{0.940}
& \textbf{0.928} & \textbf{0.918} \\
\bottomrule
\end{tabular}%
}
\end{table}

\begin{table}[t]
\centering
\small
\caption{\textbf{Independent cross-evaluation of localized CDR redesign.}}
\label{tab:cdr_cross_eval}
\resizebox{0.8\linewidth}{!}{%
\begin{tabular}{lcccc|cccc}
\toprule
& \multicolumn{4}{c|}{AF2-IG}
& \multicolumn{4}{c}{Rosetta} \\
Method
& iPAE $\downarrow$ & iPTM $\uparrow$ & pTM $\uparrow$ & pLDDT $\uparrow$
& $\Delta G \downarrow$ & SC $\uparrow$ & H-bonds $\uparrow$ & Packstat $\uparrow$ \\
\midrule
RFdiffusion v3 & 17.483 & 0.505 & 0.665 & \textbf{0.871} & -103.9 & 0.681 & 14.200 & 0.488 \\
% BoltzGen & {\textbf{12.374}} & {\textbf{0.752}} & {\textbf{0.761}} & {0.842} & {-122.2} & {0.652} & {21.325} & {0.661} \\ 
BindCraft & 23.294 & 0.360 & \ 0.602 & 0.642 & -78.3 & 0.571 & 14.750 & 0.618 \\ \midrule 
% BindCraft v2 & \textcolor{blue}{16.888} & \textcolor{blue}{0.738} & \textcolor{blue}{0.657} & \textcolor{blue}{0.845} & \textcolor{blue}{-130.6} & \textcolor{blue}{0.651} & \textcolor{blue}{19.800} & \textcolor{blue}{0.664} \\ \midrule 
\textbf{ProteoCraft-MCTS} & \textbf{14.558} & \textbf{0.753} & \textbf{0.716} & 0.851 & \textbf{-138.3} & \textbf{0.702} & \textbf{24.575} & \textbf{0.673} \\ \bottomrule
\end{tabular}}
\end{table}

\subsection{Prospective Experimental Validation of Designed Binders}

To assess whether ProteoCraft designs exhibit experimentally measurable binding, we prospectively evaluated de novo miniprotein binders against programmed death-ligand 1 (PD-L1) and VHH nanobody designs against interferon alpha/beta receptor 2 (IFNAR2), two therapeutically relevant cell-surface targets. Candidate sequences were selected using the complete ProteoCraft pipeline, integrating ESMFold2 complex confidence, independent structural cross-evaluation, and interface-quality filtering. The selected proteins were synthesized and tested for direct target binding using surface plasmon resonance (SPR) for PD-L1 and biolayer interferometry (BLI) for IFNAR2.

\begin{figure}[t]
    \centering
    \includegraphics[width=0.49\linewidth]{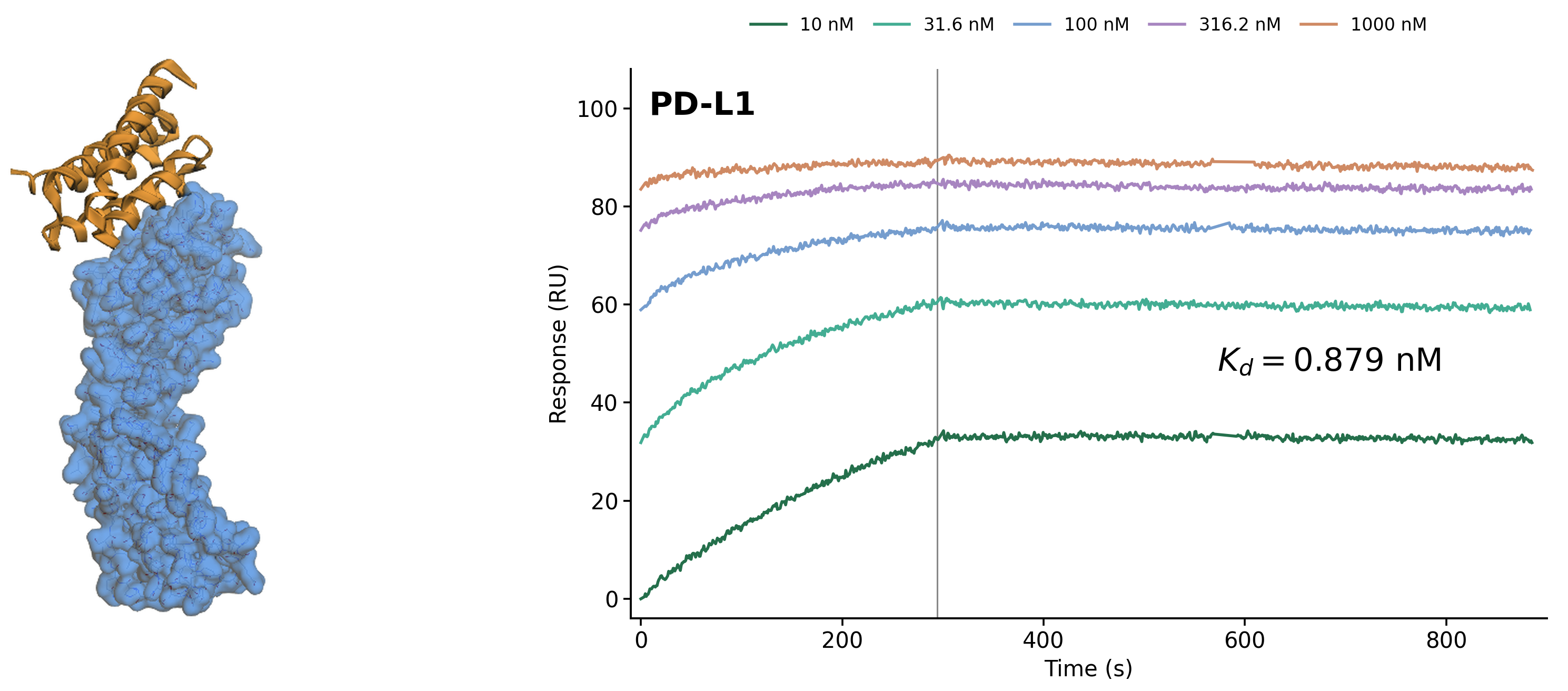}
    \hfill
    \includegraphics[width=0.49\linewidth]{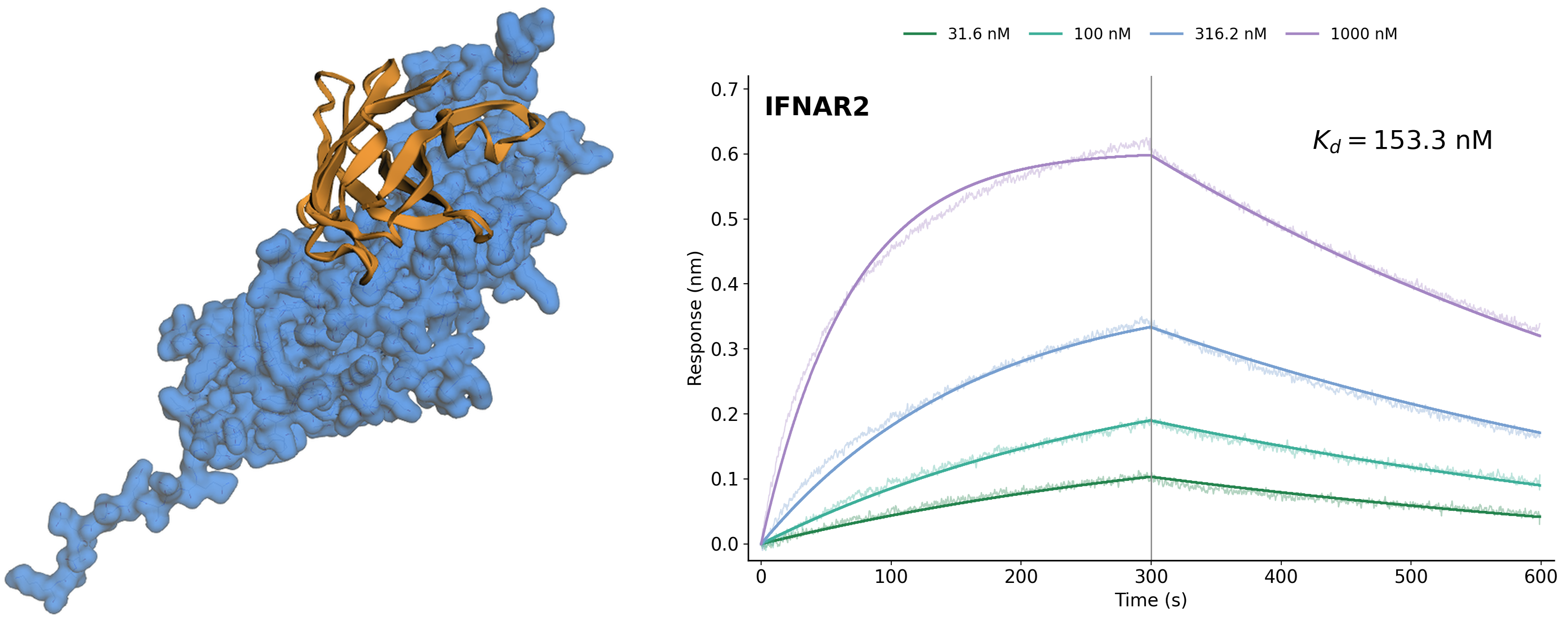}
    \caption{\textbf{Wet-lab validation of ProteoCraft-designed binders.}
    Complex structures and biolayer interferometry sensorgrams for the selected
    de novo binders against PD-L1 (left) and IFNAR2 (right).}
    \label{fig:wetlab_validation}
    \vspace{-1em}
\end{figure}

The PD-L1 miniprotein group achieved the highest binding hit rate among the evaluated groups. All 35 candidates met the experimental expression/loading criteria and proceeded to binding measurements. Of these, 29 were classified as binding hits under the assay-specific criteria, corresponding to a hit rate of 82.9\%. Based on the selected kinetic fits in the latest experimental report, six candidates exhibited $K_D < 10$\,nM (6/35; 17.1\%): five had $1 \leq K_D < 10$\,nM, and one achieved subnanomolar affinity. The strongest candidate yielded a fitted $K_D=0.879$\,nM in a single SPR measurement (Fig.~\ref{fig:wetlab_validation}).

For IFNAR2, four of the nine VHH candidates that met the experimental expression criteria were classified as binding hits, corresponding to a hit rate of 44.4\% among successfully expressed candidates. The representative lead exhibited concentration-dependent binding in BLI measurements, with fitted $K_D$ values of 153.3 and 160.3\,nM in two independent measurements. These results extend the experimental validation beyond miniproteins to an immunoglobulin-like scaffold, where the designed antigen-binding regions must remain compatible with the VHH framework.

These binders were obtained from the computational design and filtering pipeline without experimental affinity maturation. Together, these results demonstrate experimentally measurable target binding across two distinct design settings, including a high binding hit rate and subnanomolar affinity in the PD-L1 miniprotein group. Further characterization will be required to establish binding specificity, biological activity, and generalization across additional targets.

\subsection{Ablation Studies and Discussion}
\label{subsec:analysis}

\paragraph{Contribution of One-step diffusion.}
Tab.~\ref{tab:denovo_esmf2} removes only the One-step diffusion gradient. Adding this signal improves mean Pair-chain iPTM from $0.637$ to $0.660$ before tree search and increases the number of candidates above Pair iPTM $0.7$ from $85$ to $118$. This supports the role of truncated coordinate-level sensitivity as a \emph{proposal signal}: it enriches the candidate set before full verification, even though the truncated score itself is never used as the final candidate value.

\paragraph{Contribution of Stage B refinement.} We assess the benefit of single-site refinement on the 14 de novo binders in Tab.~\ref{tab:denovo_esmf2}. For each setting, both branches resume from the same checkpoint after 70 Stage A steps and receive 40 additional search steps, either continuing Stage A multi-site exploration or switching to Stage B single-site refinement. Stage B increases mean Pair-chain iPTM from $0.691$ to $0.721$. It also improves all other reported ESMFold2 confidence metrics on average. This supports the benefit of Stage B refinement over extending Stage A under this evaluation setting.

\begin{figure}[t]
    \centering
    \includegraphics[width=0.98\linewidth]{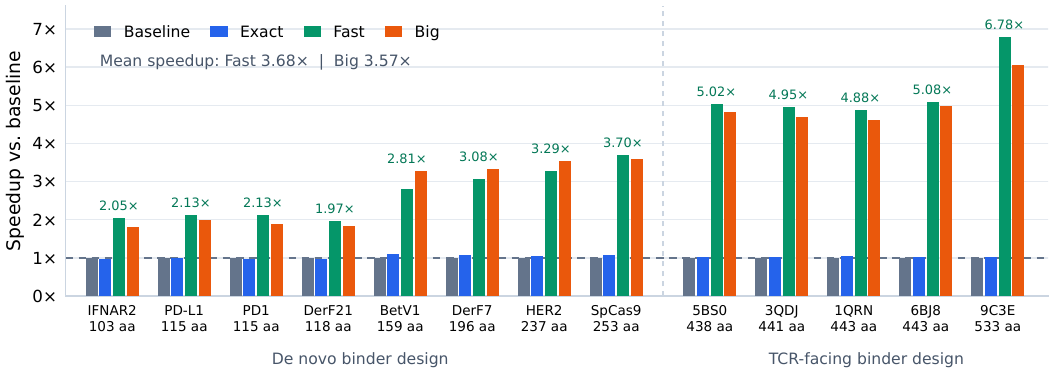}
    \vspace{-0.5em}
    \caption{\textbf{Accelerating ProteoCraft through inference optimization.}}
    % \vspace{-1.5em}
    \label{fig:speedup}
\end{figure}
\newpage
\begin{wraptable}{r}{0.36\textwidth}
\centering
\small
\vspace{-1.5em}
\setlength{\tabcolsep}{5pt}
\renewcommand{\arraystretch}{1.15}
\caption{\textbf{Runtime comparison.} We report mean wall-clock latency (seconds) per binder under each method's native design and selection procedure. Details are in Appendix~\ref{app:runtime_protocol}.}
\label{tab:runtime_comparison}
\resizebox{\linewidth}{!}{%
\begin{tabular}{lcc}
\toprule
Method & De novo & TCR \\
\midrule
PXDesign-d        & 18.72   & 21.01 \\
Proteina-Complexa & 14.62   & 34.61 \\
RFdiffusion v3   & 56.64   & 44.26 \\
BoltzDesign1     & 693.38  & 2882.20 \\
BindCraft v2     & 2321.50 & --- \\
\midrule
ProteoCraft      & 32.34   & 174.06 \\
\bottomrule
\end{tabular}}
\vspace{-2.5em}
\end{wraptable}
\paragraph{Runtime and speedup.}
Table~\ref{tab:runtime_comparison} summarizes per-binder computational costs under method-specific output definitions. ProteoCraft requires an average of 32.34 seconds per qualifying candidate on de novo targets and 174.06 seconds on TCR targets, counting unique search candidates with overall iPTM $>0.70$.  We further evaluate whether inference optimization can accelerate the full ProteoCraft search pipeline using the ESMFold2 optimization kit released by Anthropic.\footnote{\url{https://github.com/anthropics/uplifting-biomolecular-modeling}}
Across 13 targets, Fast and Big achieve mean end-to-end speedups of $3.68\times$ and $3.57\times$ over unoptimized ProteoCraft, respectively, with Fast reaching $6.78\times$ on 9C3E (Fig.~\ref{fig:speedup}). These results demonstrate that inference-level improvements can substantially reduce the computational cost of verifier-guided search without performance sacrifice (Fig.~\ref{fig:proteocraft_quality}).

%% file: sec/2_related.tex
\section{Related Work}
\label{sec:related_work}
% \paragraph{ Protein Binder Design.}  
% Recent advances in generative binder design have been accelerated by methods like RFdiffusion~\citep{butcher2025novo}, ProteinMPNN~\citep{dauparas2022robust}, and AlphaProteo~\citep{zambaldi2024novo}, which combine sequence design with structure prediction. Methods like AlphaProteo~\citep{zambaldi2024novo}, Protenix~\citep{team2025pxdesign}, PXDesign~\citep{team2025pxdesign}, and BoltzGen~\citep{stark2025boltzgen} utilize pre-trained structure predictors to generate optimized sequences for binders. However, these approaches primarily rely on generative priors and are limited by the alignment between the target and training data. To improve design accuracy for complex interfaces and target-specific optimization, new approaches have begun to incorporate structure-predictor-guided search.

\paragraph{Hallucination-based Protein Design.}
Neural structure predictors such as AF2 have been repurposed as differentiable design oracles, which optimize protein sequences directly against structural objectives~\citep{anishchenko2021novo,pacesa2025one}. BindEnergyCraft introduces a denser energy-based objective for this optimization~\citep{nori2025bindenergycraft}, whereas mBER incorporates sequence and structural conditioning to support format-specific antibody design~\citep{swanson2025mber}. Extending this paradigm to AF3-style predictors~\citep{protenix2026protenix,passaro2025boltz} is challenging because their iterative diffusion-based inference makes end-to-end differentiation through the complete sampling trajectory computationally expensive. BoltzDesign1~\citep{cho2025boltzdesign1} addresses this difficulty by optimizing through the intermediate distogram without gradients from the diffusion sampling. ProteoCraft, instead, differentiates through only the terminal diffusion step and uses truncated gradients as mutation proposals.

\paragraph{Inference-Time Search and Verification.} Several methods refine protein designs through repeated generation and evaluation. EvoPro~\citep{goudy2023silico} applies a genetic algorithm in which discrete sequence variants are evaluated through complete structure-prediction runs. PPIFlow~\citep{yu2026high} couples flow-based generation with an in silico maturation stage, while Proteína-Complexa~\citep{didi2026scaling} scales inference-time compute through best-of-$N$ sampling, beam search, Feynman--Kac steering, and MCTS over generative trajectories. Search-based inference-time scaling has also been applied to all-atom diffusion models, where inexpensive intermediate rewards guide the selection of denoising trajectories~\citep{yoo2026search}.
However, existing methods either discard local gradient information or rely on approximate intermediate rewards. ProteoCraft instead uses truncated gradients to propose discrete mutations and leverages coarse-to-fine search that progresses from coordinated multi-site exploration to single-site refinement, focusing expensive evaluations on promising sequence regions.

%% file: sec/X_appendix.tex
\section{Method Details}
\label{app:implementation}
Algorithm~\ref{alg:app_proteocraft} summarizes the complete procedure.

\begin{algorithm}[h]
\caption{ProteoCraft proposal--verification search.}
\label{alg:app_proteocraft}
\begin{algorithmic}[1]
\Require Fixed context $\mathcal T$, mutable mask $\mathbf M$, initial logits
$\{\boldsymbol\Theta_r\}_{r=1}^{N_{\mathrm{root}}}$, frozen ESMFold2 model
\Ensure Best fully verified sequence and structure
\State initialize an empty global verified archive
\For{$r=1,\ldots,N_{\mathrm{root}}$}
  \State $\mathbf s_r\gets\arg\max\boldsymbol\Theta_r$; fully verify
  $\mathbf s_r$ and initialize root $\mathbf s_r^0$
  \State add $\mathbf s_r^0$ to the verified archive
\EndFor
\Statex \textbf{Stage A: multi-site exploration}
\For{$t=1,\ldots,B_A$}
  \State select verified sequence $\mathbf s^i$ by Eq.~\ref{eq:app_ucb}
  \State optimize its logits $\boldsymbol\Theta^i$ using
  Eq.~\ref{eq:app_gradient_combine}
  \State compute $\boldsymbol\Delta_{s_i,s_i'}$ and construct multi-site candidates
  \State fully verify at most eight unique candidates; archive all results
  \State insert at most two selected children and back up their rewards
\EndFor
\State choose the two best archived states and re-verify them with a common seed
\Statex \textbf{Stage B: single-site refinement}
\For{$t=1,\ldots,B_B$}
  \State select $(\mathbf s^i,a_{\mathbf s^i})$ by Eq.~\ref{eq:app_puct}
  \State apply $a_{\mathbf s^i}$ to obtain a one-residue mutant $\mathbf s'$
  \State fully verify $\mathbf s'$; archive it and add it as a child
  \State back up state and action values using Eq.~\ref{eq:app_backup}
\EndFor
\State \Return the archived $\mathbf s$ with minimum
$\mathcal L_{\mathrm{verify}}(\mathbf s)$
\end{algorithmic}
\end{algorithm}

\subsection{State Representation and Search Statistics}

Let $\mathcal T$ denote the fixed molecular context and $\mathbf s=(s_1,\ldots,s_L)$ the designed sequence. A mask $\mathbf M\in\{0,1\}^{L}$ identifies mutable positions. Each sequence is associated with logits $\boldsymbol\Theta\in\mathbb R^{L\times20}$ and soft residue probabilities
\begin{equation}
P_{\boldsymbol\Theta}(i,a)
=\operatorname{Softmax}(\boldsymbol\Theta_i/T)_a,
\qquad
s_i=\arg\max_a\boldsymbol\Theta_{i,a},
\label{eq:app_soft_sequence}
\end{equation}
where $T$ is annealed during proposal optimization. Fixed positions and target residues are represented by one-hot identities and never receive logit updates.

A verified node is represented by the same discrete sequence notation used in the main text. For node $\mathbf s^i$, we store
\begin{equation}
(\mathbf s^i,\boldsymbol\Theta^i,\mathbf X^i,
\mathcal V^i,N(\mathbf s^i),q(\mathbf s^i)),
\end{equation}
where $\mathbf X^i$ and $\mathcal V^i$ are the structure and metrics returned by complete ESMFold2 inference, $N(\mathbf s^i)$ is the visit count, and $q(\mathbf s^i)$ is the mean backed-up reward. In Stage B, a single-site action $a_{\mathbf s^i}$ additionally stores $N(\mathbf s^i,a_{\mathbf s^i})$ and $q(\mathbf s^i,a_{\mathbf s^i})$. All state and action values are therefore empirical quantities derived only from fully verified sequences.

\subsection{Differentiable Mutation Proposals}

\paragraph{Soft sequence conditioning.} The probability matrix in Eq.~\ref{eq:app_soft_sequence} is passed through the ESMFold2 \texttt{res\_type\_soft} interface. The current hard sequence defines a valid residue and atom topology, while the soft representation conditions the trunk and exposes gradients with respect to $\boldsymbol\Theta$. No straight-through estimator is required.

\paragraph{Distogram objective.} Let $p_{ijr}$ be the predicted probability that residues $i$ and $j$ occupy distance bin $r$. For cutoff $d$, we compute a differentiable contact penalty from the probability mass below $d$,
\begin{equation}
\ell_{ij}^{\mathrm{contact}}(d)
=-\log\!\left(\sum_{r:\,d_r\le d}p_{ijr}+\varepsilon\right).
\label{eq:app_contact}
\end{equation}
Aggregating Eq.~\ref{eq:app_contact} over binder--binder, target--binder, and global binder residue pairs gives $\mathcal L_{\mathrm{intra}}$, $\mathcal L_{\mathrm{inter}}$, and $\mathcal L_{\mathrm{glob}}$, respectively. The geometric proposal loss is
\begin{equation}
\mathcal L_{\mathrm{disto}}
=0.5\mathcal L_{\mathrm{intra}}
+0.5\mathcal L_{\mathrm{inter}}
+0.2\mathcal L_{\mathrm{glob}}.
\label{eq:app_disto}
\end{equation}

\paragraph{Sequence prior.} Sequence plausibility is regularized with an ESMC masked language-model loss over mutable positions $\mathcal I_M=\{i:M_i=1\}$. Let $\Omega_1,\ldots,\Omega_K\subseteq\mathcal I_M$ be sampled nonempty mask sets, and let $p_{\mathrm{ESMC}}^{(k)}(i,a)$ denote the predicted probability of amino acid $a$ at position $i$ under mask set $\Omega_k$. Using the current soft amino-acid probabilities as targets, we compute
\begin{equation}
\mathcal L_{\mathrm{ESMC}}
=
-\frac{1}{K}\sum_{k=1}^{K}
\frac{1}{|\Omega_k|}
\sum_{i\in\Omega_k}\sum_{a=1}^{20}
P_{\boldsymbol\Theta}(i,a)
\log p_{\mathrm{ESMC}}^{(k)}(i,a).
\label{eq:app_esmc}
\end{equation}
The loss is averaged over sampled mask sets and affects proposal optimization only; it is not used to verify or rank a completed candidate.

\paragraph{Truncated one-step diffusion.} The distogram supplies pairwise geometric information but does not directly differentiate through coordinate generation. We therefore retain gradients through one terminal denoising operation. Starting from $\mathbf X_{\sigma_{\max}}=\sigma_{\max}\boldsymbol\epsilon$ with $\boldsymbol\epsilon\sim\mathcal N(\mathbf0,\mathbf I)$, the reverse trajectory is run without gradient tracking to noise level $\tau$:

\begin{align}
\mathbf X_\tau
&=\operatorname{sg}\!\left[
f_\phi^{\sigma_{\max}\rightarrow\tau}
(\mathbf X_{\sigma_{\max}};\mathbf c_{\boldsymbol\Theta})\right],\\
\widehat{\mathbf X}_0
&=D_\phi(\mathbf X_\tau,\tau;\mathbf c_{\boldsymbol\Theta}),
\label{eq:app_one_step}
\end{align}

where $\operatorname{sg}$ denotes stop-gradient and $D_\phi$ is the frozen denoiser that predicts clean coordinates at noise level $\tau$. The coordinate-sensitive loss is

\begin{align}
\mathcal L_{\mathrm{1step}}={}&
0.1\mathcal L_{\mathrm{pLDDT}}
+0.4\mathcal L_{\mathrm{PAE}}
+0.2\mathcal L_{\mathrm{iPAE}}
+0.3\mathcal L_{\mathrm{pair\text{-}chain\ iPTM}}
\nonumber\\[-2pt]
&+0.1\mathcal L_{R_g}
+0.1\mathcal L_{\mathrm{clash}}
+0.1\mathcal L_{\mathrm{i\text{-}clash}},
\label{eq:app_onestep_loss}
\end{align}

with all components written as losses. In particular, $\mathcal L_{\mathrm{pair\text{-}chain\ iPTM}} =1-\mathrm{pair\text{-}chain\ iPTM}$; the PAE terms measure global and interface alignment error, and the remaining terms penalize low confidence, excessive radius of gyration, and internal or interface clashes. Gradients are retained through the terminal denoising operation and the current sequence conditioning, while the preceding reverse-diffusion trajectory is detached. The structure produced by this truncated calculation is not treated as a verified candidate.

\paragraph{Gradient combination and soft optimization.} For $\mu\in\{\mathrm{disto},\mathrm{ESMC},\mathrm{1step}\}$, let $\mathbf g_\mu=\mathbf M\odot\nabla_{\boldsymbol\Theta}\mathcal L_\mu$. We normalize each signal over active entries, $\operatorname{Norm}(\mathbf g)=\mathbf g/(\|\mathbf g\|_2+\varepsilon)$, and combine them as
\begin{equation}
\mathbf g=
\operatorname{Norm}(\mathbf g_{\mathrm{disto}})
+\lambda_{\mathrm{ESMC}}
 \operatorname{Norm}(\mathbf g_{\mathrm{ESMC}})
+\lambda_{\mathrm{1step}}
 \operatorname{Norm}(\mathbf g_{\mathrm{1step}}).
\label{eq:app_gradient_combine}
\end{equation}
At each soft step, logits are updated by $\boldsymbol\Theta\leftarrow\boldsymbol\Theta-\eta(T)\mathbf g$, with the learning rate scaled by the current temperature. Stage A and Stage B perform 30 and 20 soft steps per expansion, respectively. Importantly, the optimized argmax sequence is only one possible proposal; acceptance always requires complete verification.

\subsection{Hotspot and Region Constraints}

For a target hotspot set $\mathcal H$ and binder residue set $\mathcal B$, we define
\begin{equation}
\mathcal L_{\mathrm{hot}}
=\frac{1}{|\mathcal H|}
\sum_{i\in\mathcal H}\min_{j\in\mathcal B}
\ell_{ij}^{\mathrm{contact}}(d_{\mathrm{hot}}),
\label{eq:app_hotspot}
\end{equation}
and replace the interface term in Eq.~\ref{eq:app_disto} by
\begin{equation}
\mathcal L_{\mathrm{inter}}' =(1-\lambda_{\mathrm{hot}})\mathcal L_{\mathrm{inter}}
+\lambda_{\mathrm{hot}}\mathcal L_{\mathrm{hot}}.
\end{equation}
Hotspot distances and coverage do not enter the ranking loss in Eq.~\ref{eq:app_verify}, but may contribute to the Stage-B search reward. Localized redesign uses the same losses on the complete complex while masking all logit gradients outside the selected region. Thus, fixed chains and framework residues remain structural context without being edited.

\subsection{Converting Logit Updates into Discrete Edits}

Let $\boldsymbol\Theta$ and $\boldsymbol\Theta'$ denote the logits before and after one proposal-optimization phase. For the substitution $s_i\rightarrow s_i'$ at position $i$, the mutation benefit is
\begin{equation}
\boldsymbol\Delta_{s_i,s_i'}=
\left(\boldsymbol\Theta'_{i,s_i'}-\boldsymbol\Theta_{i,s_i'}\right)
-\left(\boldsymbol\Theta'_{i,s_i}-\boldsymbol\Theta_{i,s_i}\right).
\label{eq:app_mutation_benefit}
\end{equation}
A positive value means that continuous optimization increasingly favors $s_i'$ over retaining $s_i$. Edits at fixed positions and self-substitutions are excluded.

\paragraph{Stage-A candidates.} Stage A constructs coordinated multi-site edits with
\begin{equation}
K_{\max}=\left\lceil\rho\sum_iM_i\right\rceil.
\label{eq:app_kmax}
\end{equation}
One candidate applies the highest-benefit substitutions at up to $K_{\max}$ distinct positions. Additional candidates sample positions and replacement residues according to their positive benefits. We also retain the raw argmax sequence of $\boldsymbol\Theta'$, which is not constrained by $K_{\max}$. Duplicate and previously evaluated sequences are removed before verification.

\paragraph{Stage-B actions.} Stage B admits only single-residue substitutions. For an action $a_{\mathbf s^i}$ from state $\mathbf s^i$, its proposal prior is
\begin{equation}
P(a_{\mathbf s^i}\mid\mathbf s^i)=
\frac{\exp(\boldsymbol\Delta_{a_{\mathbf s^i}})}
{\sum_{a'_{\mathbf s^i}\in\mathcal A(\mathbf s^i)}
\exp(\boldsymbol\Delta_{a'_{\mathbf s^i}})},
\label{eq:app_mutation_prior}
\end{equation}
where $\mathcal A(\mathbf s^i)$ contains admissible, unevaluated edits from $\mathbf s^i$. The prior expresses local gradient preference; it is not an estimate of final candidate quality.

\paragraph{Logit inheritance.} A child inherits $\boldsymbol\Theta'$ from the proposal that generated it. If the selected discrete child differs from its argmax at an edited position, we minimally align that row:
\begin{equation}
\boldsymbol\Theta^{\mathrm{child}}_{i,s_i^{\mathrm{child}}}
=\max_a\boldsymbol\Theta'_{i,a}+\delta,
\qquad\delta>0,
\label{eq:app_logit_alignment}
\end{equation}
leaving all other entries unchanged. This preserves the optimized continuous state while ensuring that the inherited logits encode the verified child sequence.

\subsection{Complete-Model Verification}

Every proposed discrete sequence is evaluated by complete frozen ESMFold2 inference. Its default ranking loss is
\begin{equation}
\mathcal L_{\mathrm{verify}}(\mathbf s^j)=
(1-\mathrm{pair\text{-}chain\ iPTM})
+0.2(1-\mathrm{binder\ pLDDT})
+0.1(1-\mathrm{pTM})
+0.1\mathcal L_{R_g\text{-}\mathrm{excess}},
\label{eq:app_verify}
\end{equation}

In Stage A, the reward is defined as $R(\mathbf s^j)=-\mathcal L_{\mathrm{verify}}(\mathbf s^j)$. Pair-chain iPTM is computed only between the designated target chains and the binder. The final term is a normalized penalty applied only when binder radius of gyration exceeds its length-dependent threshold. Additional confidence, PAE, contact, coordinate, clash, and hotspot statistics are recorded separately and do not add further terms to the ranking loss in Eq.~\ref{eq:app_verify}.

Stage B uses a separate tree-search reward that balances interface confidence, binder confidence, global confidence, and binder compactness. Additional penalties account for unmet hotspot constraints, when specified, and interfaces below a quality floor. This reward guides which sequence branches are explored further, while the recorded verification loss is used to select candidates for the frozen benchmark cohort.

\subsection{Coarse-to-Fine Tree Search}

\paragraph{Stage A: multi-site basin exploration.} We initialize $N_{\mathrm{root}}=8$ independent sequence--logit roots and fully verify their argmax sequences. Each root is expanded at least twice before allocating the remaining budget adaptively. Selection uses
\begin{equation}
U(\mathbf s^i)=q(\mathbf s^i)+c_{\mathrm{ucb}}
\sqrt{\frac{\log N_{\mathrm{total}}}{N(\mathbf s^i)+\varepsilon}},
\label{eq:app_ucb}
\end{equation}
where $q(\mathbf s^i)$ is the mean backed-up reward. The same criterion is applied at the root level using root-specific visits, yielding hierarchical allocation across initializations and then within each tree. After selection, the inherited logits are optimized, multi-site candidates are constructed and fully verified, and the resulting reward is backed up.

\paragraph{Reward backup.} For a newly verified child $\mathbf s^j$ selected for tree insertion, the trajectory $\mathbf s^0\oplus\cdots\oplus\mathbf s^j$ is traversed toward the root. At ancestor $\mathbf s^i$, we update
\begin{equation}
N(\mathbf s^i)\leftarrow N(\mathbf s^i)+1,
\qquad
q(\mathbf s^i)\leftarrow
q(\mathbf s^i)+
\frac{R(\mathbf s^j)-q(\mathbf s^i)}{N(\mathbf s^i)},
\label{eq:app_backup}
\end{equation}
We use undiscounted backup, corresponding to $\gamma=1$. Stage-B action statistics are updated analogously for every traversed $(\mathbf s^i,a_{\mathbf s^i})$. Larger $q$ indicates a higher mean backed-up reward under the reward definition used by the current search stage.

\paragraph{Verified handoff.} After Stage A, the two best verified sequence--logit states in the archive are selected. Since ESMFold2 inference is stochastic, both are re-evaluated with a common verifier seed before becoming Stage-B roots. This prevents differences in verifier sampling from being mistaken for improvement during refinement.

\paragraph{Stage B: single-site refinement.} At state $\mathbf s^i$, action selection maximizes
\begin{equation}
U_{\mathrm{PUCT}}(\mathbf s^i,a_{\mathbf s^i})
=q(\mathbf s^i,a_{\mathbf s^i})
+c_{\mathrm{puct}}P(a_{\mathbf s^i}\mid\mathbf s^i)
\frac{\sqrt{N(\mathbf s^i)}}
{1+N(\mathbf s^i,a_{\mathbf s^i})}.
\label{eq:app_puct}
\end{equation}
The first term exploits mutations whose verified descendants have high reward; the second favors gradient-preferred actions that have received little search budget. After selecting an action, ProteoCraft creates the corresponding single-site mutant, runs complete verification, adds the verified child, and updates node and edge statistics using Eq.~\ref{eq:app_backup}.

\begin{table}[t]
\centering
\small
\caption{Default ProteoCraft--ESMFold2 configuration.}
\label{tab:default_config}
\begin{tabular}{lll}
\toprule
Parameter & Value & Description \\
\midrule
Stage-A outer steps & 70 & Multi-site search budget \\
Stage-A soft steps & 30 & Proposal updates per expansion \\
Stage-B outer steps & 40 & Single-site search budget \\
Stage-B soft steps & 20 & Proposal updates per expansion \\
Sequence temperature & $1.0\rightarrow0.01$ & Cosine annealing \\
Base learning rate & 0.05 & Scaled by current temperature \\
One-step noise level $\tau$ & 2.0 & Diffusion truncation level \\
$\lambda_{\mathrm{1step}}$ & 0.25 & Normalized gradient weight \\
Discount factor $\gamma$ & 1 & Tree-backup discount \\
Stage-A mutation ratio $\rho$ & 0.5 & Maximum edited fraction \\
Independent Stage-A roots & 8 & Initial sequence--logit states \\
Stage-A handoff states & 2 & Roots transferred to Stage B \\
Candidates per expansion & $\leq8$ & Complete verifier calls \\
Inserted children per expansion & $\leq2$ & Tree branching limit \\
\bottomrule
\end{tabular}
\end{table}

\subsection{Task Instantiations and Configuration}

Unless otherwise specified, all task instantiations use the default search configuration in Table~\ref{tab:default_config}. Task-specific hotspot definitions and coefficients are fixed within each matched comparison.

\paragraph{De novo binder design.} All designated binder positions are mutable; target chains remain fixed. Interface losses and pair-chain iPTM are evaluated between the binder and the designated target chains.

\paragraph{TCR-facing binder design.} The receptor complex is structural context, but only explicitly designated chains contribute to target--binder objectives, pair-chain iPTM, and hotspot masks.

\paragraph{Localized CDR redesign.} Only residues selected by $\mathbf M$ are updated. The remainder of the antibody or TCR framework participates in complete prediction but remains sequence-fixed. Because ESMFold2 does not consume an external template, this setting preserves the framework sequence rather than clamping input backbone coordinates.

\section{Experimental Details}
\label{app:experimental_details}

This section provides the experimental settings required to interpret and reproduce the supplementary results. We provide benchmark composition, common evaluation protocols, per-target results, independent cross-evaluation, and robustness analyses.

\subsection{Experimental Setup}
\label{app:experimental_setup}

\paragraph{Matched \emph{de novo} binder benchmark.} We evaluate 14 target settings: BetV1, DerF21-B10, DerF21-B28, DerF21-B63, DerF21-B81, DerF21-B92, DerF7-B132, DerF7-B178, HER2, HER2-B139, IFNAR2, PD-L1, PD1, and SpCas9. For each method, we generate five candidates under each of three matched binder-length settings, yielding 15 candidates per target and $14\times15=210$ candidates per method. All aggregate statistics assign equal weight to each candidate. Because different model families expose incomparable native objectives, method-specific internal losses are never compared directly.  Hotspot annotations are listed in Table~\ref{tab:type1_hotspots}. 

\paragraph{ProteoCraft variants.} \textbf{No One-step Diffusion} removes only the truncated one-step diffusion gradient from Eq.~\ref{eq:app_gradient_combine}, retaining the distogram signal, ESMC prior, candidate construction, and full verification. \textbf{ProteoCraft-Base / Best-of-$N$} uses the same differentiable proposal mechanism and verifier budget as the full method but generates independent proposal trajectories without tree-search control. \textbf{ProteoCraft-MCTS} denotes the complete coarse-to-fine search with Stage-A multi-site exploration and Stage-B single-site PUCT refinement. These variants use identical targets, candidate counts, and downstream evaluators.

\paragraph{External baselines.} PXDesign-d, RFdiffusion v3, BoltzDesign1, and BindCraft are run from the same target definitions whenever supported by the corresponding method. Candidate counts are matched within each reported comparison. We use each method's native generation procedure, but discard native ranking scores during cross-method evaluation; all retained sequences are re-evaluated by the common protocols below.

\paragraph{TCR-facing binder design.} We additionally design binders against five immune-relevant TCR surfaces (Table~\ref{tab:tcr_targets}). The target TCR remains fixed and the new binder is generated from scratch. The matched comparison contains $N=42$ designs per method.

\begin{table}[t]
\centering
\scriptsize
\caption{Hotspot definitions for eight de novo target proteins and five TCR-facing target systems.}
\label{tab:type1_hotspots}
\begin{tabular}{@{}p{0.09\linewidth}p{0.21\linewidth}p{0.60\linewidth}@{}}
\toprule
Case & Group & Hotspot residues \\
\midrule
BetV1 & Normal binder & B24 \\
DerF21 & Normal binder & B28 \\
DerF7 & Normal binder & B132 \\
HER2 & Normal binder & B134, B139 \\
IFNAR2 & Normal binder & B45, B73, B75, B77, B89, B91 \\
PD-L1 & Normal binder & B37, B39, B49, B98 \\
PD1 & Normal binder & B33, B95, B98, B102 \\
SpCas9 & Normal binder & B134 \\
1QRN & \mbox{TCR-target binder} & A19--A23; A53--A56; A63--A66; A72--A76; B21--B24; B5--B7; B66--B69; B74--B77 \\
3QDJ & \mbox{TCR-target binder} & A53--A56; A61--A64; A70--A74; B23--B27; B68--B69; B78--B80; B7--B10 \\
5BS0 & \mbox{TCR-target binder} & A20--A25; A57--A59; A63--A67; A73--A77; B20--B24; B66--B69; B75--B78; B7--B8 \\
6BJ8 & \mbox{TCR-target binder} & A20--A24; A64--A68; A73--A77; B20--B24; B4--B7; B65--B68; B73--B77 \\
9C3E & \mbox{TCR-target binder} & A75--A77; A82--A85; A91--A94; B24--B26; B39--B43; B84--B85; B92--B96 \\
\bottomrule
\end{tabular}
\end{table}

\begin{table}[t]
\centering
\small
\caption{TCR-facing binder-design systems.}
\label{tab:tcr_targets}
\begin{tabular}{lll}
\toprule
PDB / case & TCR target & Antigen / restriction \\
\midrule
1QRN & A6 TCR & TAX-A2 \\
3QDJ & DMF5 TCR & MART1-A2 \\
5BS0 & A3A TCR & MAGEA3-A1 / Titin-A1 \\
6BJ8 & TCR55 & HIV-B35 \\
9C3E & 1G4 TCR & NYESO-A2 \\
\bottomrule
\end{tabular}
\end{table}

\paragraph{Localized CDR redesign.} We evaluate four chain-specific redesign settings from two scaffolded TCR complexes (Table~\ref{tab:cdr_settings}). Only the indicated CDR3 residues are mutable; the peptide, MHC, partner TCR chain, and fixed residues of the redesigned chain remain in the modeled context. We evaluate ten candidates per case for RFdiffusion v3, BindCraft, and ProteoCraft-MCTS, yielding $N=40$ candidates per method.

\begin{table}[t]
\centering
\small
\caption{Localized CDR redesign settings.}
\label{tab:cdr_settings}
\begin{tabular}{lll}
\toprule
Case & CDR3 sequence & Residue range \\
\midrule
9NMW $\alpha$ & CAVNNARLMFGDGTQL & 90--105 \\
9NMW $\beta$  & CASSVAGSPEAFF & 89--101 \\
1QRN $\alpha$ & CAVTTDSWGKLQF & 88--100 \\
1QRN $\beta$  & CASRPGLAGGRPEQYF & 91--106 \\
\bottomrule
\end{tabular}
\end{table}

\subsection{Common Evaluation Protocols}
\label{app:evaluation_protocol}

\paragraph{Complete ESMFold2 evaluation.} Every final discrete sequence is independently re-evaluated by complete frozen ESMFold2 inference. We report Pair-chain iPTM for the designated target--binder pair, complex iPTM, pTM, complex pLDDT, and pLDDT restricted to the designed chain. For localized redesign, Peptide Pair iPTM measures the redesigned chain against the peptide, while Mean Pair iPTM averages its pairwise confidence against all fixed context chains. The verifier loss is defined in Eq.~\ref{eq:app_verify}. Threshold yields are used only as reporting statistics and do not replace continuous metrics.

\paragraph{Independent AF2-IG evaluation.} AF2-IG independently predicts each final sequence and is never used for ProteoCraft proposal generation, tree selection, or reward assignment. We report interface predicted alignment error (iPAE; lower is better), iPTM, pTM, and pLDDT where available. Because AF2-IG is outside the search loop, agreement between ESMFold2 and AF2-IG provides evidence that improvements are not restricted to the search-time verifier.

\paragraph{Rosetta interface evaluation.} All methods use the same post-generation Rosetta pipeline. Each candidate is converted to a common manifest containing its method, case, binder sequence, target structure, binder length, and segment metadata, followed by standardized side-chain refinement and interface analysis. We report interface energy $\Delta G$ (more negative is better), buried solvent-accessible surface area $\Delta\mathrm{SASA}$, Lawrence--Colman shape complementarity (SC), $\Delta G/\Delta\mathrm{SASA}$, interface hydrogen bonds, and Packstat. Rosetta does not participate in sequence generation or tree search.

\subsection{Supplementary Results}
\label{app:denovo_supplementary}

\paragraph{Per-target ESMFold2 interface confidence.} Table~\ref{tab:per_target_pairiptm} reports mean Pair-chain iPTM over the 15 candidates for every target. The ProteoCraft-MCTS column uses the final frozen 210-candidate cohort, independently evaluated with ESMFold2 using 20 recycling loops and 200 diffusion sampling steps. Among the methods listed in the table, ProteoCraft-MCTS obtains the highest mean on 13 of 14 settings; Best-of-$N$ achieves the highest mean on SpCas9.
\begin{table}[t]
\centering
\scriptsize
\caption{Mean ESMFold2 Pair-chain iPTM across the 15 candidates for each matched \emph{de novo} target. Higher is better.}
\label{tab:per_target_pairiptm}
\resizebox{\linewidth}{!}{
\begin{tabular}{lcccccc}
\toprule
Target & PXDesign-d & RFdiffusion v3 & BoltzDesign1 & No One-step & Base / Best-of-$N$ & ProteoCraft-MCTS \\
\midrule
BetV1       & 0.257 & 0.185 & 0.087 & 0.793 & 0.725 & \textbf{0.832} \\
DerF21-B10  & 0.462 & 0.241 & 0.214 & 0.698 & 0.746 & \textbf{0.771} \\
DerF21-B28  & 0.449 & 0.170 & 0.099 & 0.630 & 0.729 & \textbf{0.770} \\
DerF21-B63  & 0.378 & 0.206 & 0.140 & 0.699 & 0.738 & \textbf{0.787} \\
DerF21-B81  & 0.392 & 0.140 & 0.178 & 0.600 & 0.725 & \textbf{0.756} \\
DerF21-B92  & 0.340 & 0.194 & 0.219 & 0.716 & 0.695 & \textbf{0.746} \\
DerF7-B132  & 0.212 & 0.202 & 0.104 & 0.618 & 0.603 & \textbf{0.679} \\
DerF7-B178  & 0.312 & 0.215 & 0.121 & 0.587 & 0.632 & \textbf{0.714} \\
HER2        & 0.415 & 0.262 & 0.137 & 0.635 & 0.648 & \textbf{0.776} \\
HER2-B139   & 0.278 & 0.307 & 0.194 & 0.682 & 0.614 & \textbf{0.787} \\
IFNAR2      & 0.182 & 0.233 & 0.109 & 0.644 & 0.675 & \textbf{0.704} \\
PD-L1       & 0.435 & 0.316 & 0.246 & 0.773 & 0.795 & \textbf{0.798} \\
PD1         & 0.328 & 0.151 & 0.224 & 0.686 & 0.655 & \textbf{0.721} \\
SpCas9      & 0.064 & 0.068 & 0.088 & 0.156 & \textbf{0.263} & 0.255 \\
\bottomrule
\end{tabular}}
\end{table}

\begin{table}[t]
\centering
\scriptsize
\caption{Runtime comparison across different targets, reported in seconds per binder. Averages are computed separately over eight de novo and five TCR cases.  Besides, --- indicates no filter-passing designs within the TCR sampling budget for BindCraft v2. }

\label{tab:runtime_comparison_per_case}
\resizebox{\linewidth}{!}{
\begin{tabular}{lccccccc}
\toprule
Case & Total Length & PXDesign-d & RFdiffusion3 & Complexa BoN+AF2 & BoltzDesign1 & BindCraft v2 & ProteoCraft \\
\midrule
\multicolumn{8}{l}{\textit{De novo binder design}} \\
IFNAR2 & 199 & 13.53 & 49.21 & 14.28 & 458 & 512 & 11.62 \\
PD-L1 & 208 & 14.13 & 49.40 & 14.10 & 504 & 303 & 13.54 \\
PD1 & 220 & 16.40 & 63.64 & 12.05 & 505 & 1128 & 14.96 \\
DerF21 & 257 & 11.49 & 56.46 & 15.05 & 522 & 253 & 17.17 \\
BetV1 & 261 & 14.40 & 48.17 & 16.05 & 686 & 11523 & 19.76 \\
DerF7 & 311 & 18.77 & 55.11 & 17.07 & 782 & 1381 & 36.97 \\
HER2 & 322 & 28.93 & 70.43 & 13.91 & 1021 & 681 & 45.23 \\
SpCas9 & 369 & 32.13 & 60.69 & 14.45 & 1069 & 2791 & 99.50 \\
\midrule
\textbf{Average} & --- & \textbf{18.72} & \textbf{56.64} & \textbf{14.62} & \textbf{693.38} & \textbf{2321.50} & \textbf{32.34} \\
\midrule
\multicolumn{8}{l}{\textit{TCR-facing binder design}} \\
5BS0 & 526 & 19.74 & 42.79 & 33.55 & 2648 & --- & 52.60 \\
3QDJ & 529 & 19.92 & 43.29 & 31.98 & 2616 & --- & 52.32 \\
1QRN & 531 & 20.13 & 46.76 & 33.58 & 2698 & --- & 54.30 \\
6BJ8 & 531 & 20.10 & 42.33 & 33.29 & 2680 & --- & 115.02 \\
9C3E & 621 & 25.17 & 46.15 & 40.67 & 3769 & --- & 596.08 \\
\midrule
\textbf{Average} & --- & \textbf{21.01} & \textbf{44.26} & \textbf{34.61} & \textbf{2882.20} & --- & \textbf{174.06} \\
\bottomrule
\end{tabular}}
\end{table}
\paragraph{Runtime Analysis}
\label{app:runtime_protocol}

Runtime measurements cover 13 shared cases, comprising eight binder design cases and five TCR cases. We measure the wall-clock time required by each method's native generation and selection procedure to produce a final binder in Table~\ref{tab:runtime_comparison_per_case}. For trajectory-based hallucination methods, the runtime includes executing the design trajectory and selecting the lowest-loss binder produced by that trajectory. For diffusion-based methods, it is the runtime to generate a binder directly. Thus, the comparison reflects the end-to-end time to obtain one final design under each method's native procedure. Specifically, BoltzDesign1 timings are per generated binder without native redesign or AF3 filtering. For BindCraft v2, timings are summed single-GPU generation durations divided by designs passing its internal filters. ProteoCraft timings are cumulative seconds divided by unique sequence candidates with search-time overall iPTM $>0.70$, including intermediate candidates, rather than only final outputs.

\begin{figure}[t]
    \centering
    \includegraphics[width=\linewidth]{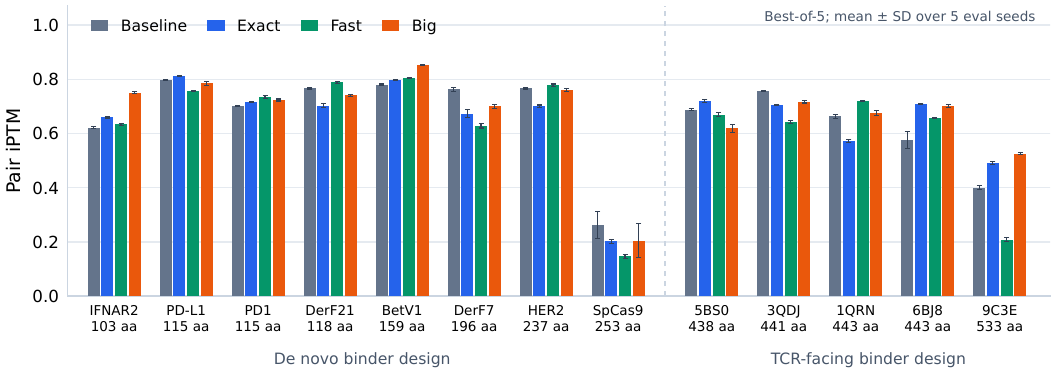}
    \caption{\textbf{Design quality with accelerated ESMFold2 inference.}}
    \label{fig:proteocraft_quality}
\end{figure}

\paragraph{Design quality after inference optimization.}
Figure~\ref{fig:proteocraft_quality} compares candidates obtained with different ESMFold2 inference configurations, all re-evaluated using unoptimized ESMFold2.
Mean pair-chain iPTM across 13 targets is 0.657 for Baseline, 0.651 for Exact, 0.628 for Fast, and 0.673 for Big.
Although individual targets exhibit both improvements and declines, aggregate scores remain broadly comparable.
In particular, Big combines a $3.57\times$ mean search speedup with a slightly higher mean pair-chain iPTM than Baseline, while Fast achieves $3.68\times$ speedup with a modest reduction.
These results suggest that inference optimization can improve search efficiency while retaining comparable average design quality, although multiple search seeds are needed to establish the robustness of this trade-off.

\begin{table}[t]
\centering
\small
\caption{\textbf{Sensitivity to the one-step diffusion noise level.}
All settings use the same three targets and matched search configuration.}
\label{tab:tau_sensitivity}
\begin{tabular*}{\linewidth}{@{\extracolsep{\fill}}c|cccc@{}}
\toprule
$\tau$ & Pair iPTM $\uparrow$ & iPTM $\uparrow$ & pTM $\uparrow$ & pLDDT $\uparrow$ \\
\midrule
1 & 0.682 & 0.871 & 0.896 & 0.886 \\
2 & \textbf{0.777} & \textbf{0.930} & \textbf{0.939} & \textbf{0.921} \\
4 & 0.730 & 0.914 & 0.931 & 0.909 \\
\bottomrule
\end{tabular*}
\end{table}
\paragraph{One-step noise level.}
The default One-step diffusion noise level is $\tau=2$.
We vary only $\tau\in\{1,2,4\}$ on three matched \emph{de novo} targets. The default $\tau=2$ gives the strongest mean result across all four metrics, while performance remains competitive at both smaller and larger noise levels.